\documentclass[12pt]{article}
\usepackage{a41} %
\usepackage{epsfig} %
\usepackage{amssymb} %
\usepackage{amsmath} %
\usepackage{verbatim} %
\usepackage{cite} %
\newcommand{\Ahat}{\hat{A}}
\newcommand{\N}{\nonumber}
\newcommand{\ep}{\varepsilon}

\newcommand{\Ctil}{\tilde{C}}

\newcommand{\beq}{\begin{equation}}
\newcommand{\eeq}{\end{equation}}
\newcommand{\bea}{\begin{eqnarray}}
\newcommand{\eea}{\end{eqnarray}}

\newcommand{\gsim}{\raisebox{-0.07cm}{$\, \stackrel{>}{{\scriptstyle
\sim}}\, $}}

\newcounter{lin}

\graphicspath{/usr1/hboett/plots/qcd_pol/}
\graphicspath{./}

\begin{document}
\begin{titlepage}

\begin{flushleft}
DESY 10--109 \hfill 
\\
TTK--10--46\\
SFB-CPP/10--077\\
August 2010 \\
\end{flushleft}

\vspace{3cm}
\noindent
\begin{center}
{\LARGE\bf \boldmath The $O(\alpha_s^3)$ Massive Operator Matrix Elements} 
\vspace*{2mm}
\noindent
{\LARGE\bf \boldmath of $O(n_f)$ for the Structure Function $F_2(x,Q^2)$}
\vspace*{2mm}
\noindent
{\LARGE\bf \boldmath and Transversity}\\
\end{center}
\begin{center}

\vspace{2cm}

{\large J. Ablinger$^a$, J. Bl\"umlein$^b$, S. Klein$^c$, C. Schneider$^a$, and F. Wi\ss{}brock$^b$}

\vspace{2cm}
{\it $^a$~Research Institute for Symbolic Computation (RISC),\\
                          Johannes Kepler University, Altenbergerstra\ss e 69,
                          A--4040, Linz, Austria}\\

\vspace*{2mm}
{\it $^b$~Deutsches Elektronen Synchrotron, DESY,\\
Platanenallee 6, D--15738 Zeuthen, Germany}\\

\vspace*{2mm}
{\it $^c$~Institut f\"ur Theoretische Teilchenphysik und Kosmologie, \\
                 RWTH Aachen University, D--52056 Aachen, Germany}

\vspace{3cm}
\end{center}

\begin{abstract}
\noindent
The contributions $\propto n_f$ to the $O(\alpha_s^3)$ massive operator matrix elements describing the
heavy flavor Wilson coefficients in the limit $Q^2 \gg m^2$ are computed for the structure function
$F_2(x,Q^2)$ and transversity for general values of the Mellin variable $N$. Here, for two matrix
elements, $A_{qq,Q}^{\sf PS}(N)$ and $A_{qg,Q}(N)$, the complete result is obtained. A first independent
computation of the contributions to the 3--loop anomalous dimensions $\gamma_{qg}(N)$, $\gamma_{qq}^{\sf PS}(N)$,
and $\gamma_{qq}^{\sf NS,(TR)}(N)$ is given. In the computation advanced summation technologies for nested
sums over products of hypergeometric terms with harmonic sums have been used. For intermediary results 
generalized harmonic sums occur, while the final results can be expressed by nested harmonic sums only.
\end{abstract}
\end{titlepage}

\newpage
\sloppy

\section{Introduction}
\label{sec:intro}

\vspace{2mm}
\noindent
The heavy flavor corrections to deep-inelastic structure functions amount to
large contributions at lower values of the Bjorken variable $x$. Currently
they are known in semi-analytic form to 2--loop (NLO) order~\cite{NLO}. The present 
accuracy of the deep-inelastic data reaches the order of 1\%~\cite{H1Z:2009wt}, which
requires the next-to-next-to-leading order (NNLO) corrections for precision determinations
of both the strong coupling constant $\alpha_s(M_Z^2)$ and the parton distribution
functions~\cite{PDF}, as well as the detailed understanding of the heavy flavor production cross
sections in lepton--nucleon scattering~\cite{EXP}.
The precise knowledge of these quantities is of central importance 
for the interpretation of the physics results at the Large Hadron Collider, LHC,~\cite{HERALHC}. 
In the region 
$Q^2 \gg m^2$, with $Q^2=-q^2$, with $q$ the space-like 4--momentum transfer and $m$ the 
heavy quark mass, the power corrections $O((m^2/Q^2)^k), k \geq 1$ to the heavy quark 
structure functions become very small. For the structure function $F_2(x,Q^2)$ the 
logarithmic and constant contributions are sufficient at the 1\%-level to describe
the complete result for $Q^2/m^2~\gsim~10$, a region which does well compare to the
deep-inelastic region at HERA in which the twist-2 contributions dominate, 
cf.~\cite{Blumlein:2008kz}.~\footnote{For higher order corrections to the gluonic contributions in the 
threshold region, cf.~\cite{Presti:2010pd}.} In 
this limit the Wilson coefficients with $n_f$ massless 
and one massive quark factorize into massive operator matrix elements (OMEs) and the
massless Wilson coefficients, as has been shown in Ref.~\cite{Buza:1995ie}. The former 
quantities are process independent, while the latter depend on the respective scattering
process. The massless Wilson coefficients for the structure function $F_2(x,Q^2)$ are
known to 3-loop order,~\cite{Vermaseren:2005qc}. 

For fixed Mellin moments $N$ a series of moments up to $N = 10 ... 14$, depending on the
respective transition, have been calculated for all the OMEs at 3--loop order contributing 
to the structure function $F_2(x,Q^2)$ and those needed to establish a variable flavor scheme 
description at $O(\alpha_s^3)$ in Ref.~\cite{Bierenbaum:2009mv}.~\footnote{For the corresponding
contributions in case of transversity see \cite{Blumlein:2009rg}.} There also the complete
renormalization of the matrix elements has been derived. In this computation the massive OMEs
for given total spin $N$ were mapped onto massive tadpoles which were computed using {\tt MATAD}, 
\cite{Steinhauser:2000ry}. For general values of $N$ the 2-loop OMEs, up to $O(\varepsilon)$, 
have been calculated in Refs.~\cite{TOL1a,TOL1b,TOL2}. All the logarithmic contributions to the massive OMEs 
are known~\cite{Bierenbaum:2010jp,BBKW} for general values of $N$. They depend on the 3-loop anomalous 
dimensions \cite{ANDIM2a,ANDIM2b}. For the structure function $F_L(x,Q^2)$ the 
asymptotic heavy flavor Wilson coefficients at $O(\alpha_s^3)$ were calculated in \cite{Blumlein:2006mh}. 
They become, however, effective only at much higher scales of $Q^2$ compared to the case of $F_2(x,Q^2)$.

In the present paper the $O(\alpha_s^3)$ contributions $\propto n_F T_F^2 C_{F,A}$ are computed
for all massive operator matrix elements contributing to the structure function $F_2(x,Q^2)$
at general values of the Mellin variable $N$ in the fixed flavor number scheme, as well as the 
corresponding contributions to transversity. This scheme has to be considered as the genuine scheme
in quantum field theoretic calculations since the initial states, the twist--2 {\it massless} partons
can, at least to a good approximation, be considered as LSZ-states. This is not the case
for heavy quark states, which have a finite lifetime.~\footnote{From the representations obtained in the 
fixed flavor number scheme, variable flavor number schemes may be defined
under specific conditions~\cite{TOL1a,Bierenbaum:2009mv} observing the correct matching 
scales~\cite{Blumlein:1998sh}.}
For two OMEs, $A_{qq,Q}^{\sf PS}(N)$ and $A_{qg,Q}(N)$, the 
complete result is obtained. In the present computation the Feynman parameter integrals are computed 
directly. They can be represented in terms of generalized hypergeometric functions \cite{HYPER} and sums 
thereof prior the expansion in the dimensional variable $\varepsilon = D -4$, 
cf.~\cite{Blumlein:2009ta,Bytev:2009kb}. 
Finally, they are represented
in terms of nested sums over products of hypergeometric terms and harmonic sums, which can be calculated 
using modern summation techniques \cite{SIGMA,HARMS} that are based on a refined difference field of 
\cite{KARR} and that generalize the summation paradigms presented in \cite{AEB} to multi-summation. 
During this computation the results can be 
expressed in terms of nested harmonic sums~\cite{HSUM1,HSUM2}. In intermediary steps of the calculation 
generalized harmonic sums, \cite{GHSUM1,GHSUM2}, cf. also~\cite{GHSUM3}, appear which finally cancel. 

The paper is organized as 
follows. In Section~\ref{sec:forma} we summarize the basic formalism. The results for the constant part 
of 
the $O(\alpha_s^3)$ $n_f$-contributions to the massive OMEs $\Ahat_{Qg}(N)$, $\Ahat_{Qq}^{\sf PS}(N)$ 
$\Ahat_{qq,Q}^{\sf NS}(N)$, $\Ahat_{qg,Q}(N)$, $\Ahat_{qq,Q}^{\sf PS}(N)$, and 
$\Ahat_{qq,Q}^{\sf NS,TR}(N)$,~cf.~\cite{Bierenbaum:2009mv,Blumlein:2009rg}, are presented in 
Section~\ref{sec:OME}. The single pole terms in $\varepsilon$ allow
to derive the terms $\propto n_f$ of the 3--loop anomalous dimensions for general values of $N$. They
are compared to the results in 
Refs.~\cite{ANDIM2a,ANDIM2b,Gracey:2003yrxGracey:2003mrxGracey:2006zrxGracey:2006ah} 
and are obtained in a first independent calculation
for $\gamma_{qg}(N)$, $\gamma_{qq}^{\sf PS}(N)$, $\gamma_{qq}^{\sf NS,TR}(N)$, in  
Section~\ref{sec:andim}.  Section~\ref{sec:concl} contains the conclusions. Some technical details of the 
calculation are given in the Appendix.
\section{The heavy flavor Wilson coefficients in the asymptotic region}
\label{sec:forma}

\noindent
The heavy flavor contributions to the structure functions $F_{(2,L)}(x,Q^2)$ with $n_f$ massless and one
heavy flavor are given by,~\cite{Bierenbaum:2009mv}~:
{\small
\begin{eqnarray}
      \label{eqF2}
       F_{(2,L)}^{Q\overline{Q}}(x,n_f+1,Q^2,m^2) &=&
       \sum_{k=1}^{n_f}e_k^2\Biggl\{
                   L_{q,(2,L)}^{\sf NS}\left(x,n_f+1,\frac{Q^2}{m^2}
                                                ,\frac{m^2}{\mu^2}\right)
                \otimes
                   \Bigl[f_k(x,\mu^2,n_f)+f_{\overline{k}}(x,\mu^2,n_f)\Bigr]
\N\\ &&\hspace{-43mm}
               +\frac{1}{n_f}\Bigl[L_{q,(2,L)}^{\sf PS}\left(x,n_f+1,\frac{Q^2}{m^2}
                                                ,\frac{m^2}{\mu^2}\right)
                \otimes
                   \Sigma(x,\mu^2,n_f)
               + L_{g,(2,L)}^{\sf S}\left(x,n_f+1,\frac{Q^2}{m^2}
                                                 ,\frac{m^2}{\mu^2}\right)
                \otimes
                   G(x,\mu^2,n_f)
                             \Bigr] \Biggr\}
\N\\ &&\hspace{-43mm}
+ e_Q^2\Biggl[
                   H_{q,(2,L)}^{\sf PS}\left(x,n_f+1,\frac{Q^2}{m^2}
                                        ,\frac{m^2}{\mu^2}\right)
                \otimes
                   \Sigma(x,\mu^2,n_f)
                  +H_{g,(2,L)}^{\sf S}\left(x,n_f+1,\frac{Q^2}{m^2}
                                           ,\frac{m^2}{\mu^2}\right)
                \otimes
                   G(x,\mu^2,n_f)
                                  \Biggr]~.\N\\
\end{eqnarray}
}

\vspace*{-3mm}\noindent
The different Wilson coefficients are denoted by $L_i, H_i$ in case the photon couples to
a light $(L)$ or the heavy $(H)$ quark, for the flavor non--singlet {\sf (NS)}, pure--singlet {\sf (PS)},
and singlet {\sf (S)} cases. Here, $\otimes$ is the Mellin convolution,
\begin{eqnarray}
[A \otimes B](x) = \int_0^1 \int_0^1 dx_1 dx_2~\delta(x - x_1 x_2) A(x_1) B(x_2)~,
\end{eqnarray}
with boundaries for the Wilson coefficients $[x(1+4m^2/Q^2), 1]$, $e_k$ the light and $e_Q$ the heavy quark 
charges. 
$\mu^2$ denotes the factorization scale, and $f_k, f_{\overline{k}}, 
\Sigma$ and $G$
are the quark, antiquark, flavor singlet and gluon momentum distribution functions, with
\begin{eqnarray}
\Sigma(x,\mu^2,n_f) = \sum_{k=1}^{n_f} \left[ f_k(x,\mu^2,n_f) + f_{\overline{k}}(x,\mu^2,n_f)\right]~.
\end{eqnarray}

For $Q^2 \gg m^2$ the massive Wilson coefficients can be expressed in terms of the renormalized
massive OMEs $A_{ij}$ and the massless Wilson coefficients $C_j$. To $O(a_s^3)$ they read ($a_s = 
\alpha_s/(4\pi))$, cf.~\cite{Bierenbaum:2009mv}~:
\begin{eqnarray}
\label{eqWIL1}
     L_{q,(2,L)}^{\sf NS}(n_f+1) &=&
     a_s^2 \Bigl[A_{qq,Q}^{(2), {\sf NS}}(n_f+1) \delta_2 +
     \hat{C}^{(2), {\sf NS}}_{q,(2,L)}(n_f)\Bigr]
     \N\\
     &+&
     a_s^3 \Bigl[A_{qq,Q}^{(3), {\sf NS}}(n_f+1) \delta_2
     +  A_{qq,Q}^{(2), {\sf NS}}(n_f+1) C_{q,(2,L)}^{(1), {\sf NS}}(n_f+1)
     + \hat{C}^{(3), {\sf NS}}_{q,(2,L)}(n_f)\Bigr]  
      \label{eqWIL2}
\nonumber\\
      L_{q,(2,L)}^{\sf PS}(n_f+1) &=&
     a_s^3 \Bigl[~A_{qq,Q}^{(3), {\sf PS}}(n_f+1)~\delta_2
     +  A_{gq,Q}^{(2)}(n_f)~~n_f\Ctil_{g,(2,L)}^{(1)}(n_f+1) 
     + n_f \hat{\Ctil}^{(3), {\sf PS}}_{q,(2,L)}(n_f)\Bigr]
         \label{eqWIL3}
\nonumber
\end{eqnarray}\begin{eqnarray}
      L_{g,(2,L)}^{\sf S}(n_f+1) &=&
     a_s^2 A_{gg,Q}^{(1)}(n_f+1)n_f \Ctil_{g,(2,L)}^{(1)}(n_f+1)
+
      a_s^3 \Bigl[~A_{qg,Q}^{(3)}(n_f+1)~\delta_2 \N\\ &&
     +  A_{gg,Q}^{(1)}(n_f+1)~~n_f\Ctil_{g,(2,L)}^{(2)}(n_f+1)
     +  A_{gg,Q}^{(2)}(n_f+1)~~n_f\Ctil_{g,(2,L)}^{(1)}(n_f+1)
     \N\\ && 
     +  ~A^{(1)}_{Qg}(n_f+1)~~n_f\Ctil_{q,(2,L)}^{(2), {\sf PS}}(n_f+1)
     + n_f \hat{\Ctil}^{(3)}_{g,(2,L)}(n_f)\Bigr]~,
\nonumber\\
     H_{q,(2,L)}^{\sf PS}(n_f+1)
     &=& a_s^2 \Bigl[~A_{Qq}^{(2), {\sf PS}}(n_f+1)~\delta_2
     +~\Ctil_{q,(2,L)}^{(2), {\sf PS}}(n_f+1)\Bigr]
+ a_s^3 \Bigl[~A_{Qq}^{(3), {\sf PS}}(n_f+1)~\delta_2 
\N\\
&&
     +~\Ctil_{q,(2,L)}^{(3), {\sf PS}}(n_f+1) 
     + A_{gq,Q}^{(2)}(n_f+1)~\Ctil_{g,(2,L)}^{(1)}(n_f+1)
     \N\\
&&
+ A_{Qq}^{(2), {\sf PS}}(n_f+1)~C_{q,(2,L)}^{(1), {\sf NS}}(n_f+1)
        \Bigr]~,       \label{eqWIL4}
\nonumber\\
     H_{g,(2,L)}^{\sf S}(n_f+1) &=& a_s \Bigl[~A_{Qg}^{(1)}(n_f+1)~\delta_2
     +~\Ctil^{(1)}_{g,(2,L)}(n_f+1) \Bigr]   + a_s^2 \Bigl[~A_{Qg}^{(2)}(n_f+1)~\delta_2
\N\\ &&
     +~A_{Qg}^{(1)}(n_f+1)~C^{(1), {\sf NS}}_{q,(2,L)}(n_f+1)
     +~A_{gg,Q}^{(1)}(n_f+1)~\Ctil^{(1)}_{g,(2,L)}(n_f+1)
     \N\\ &&
     +~\Ctil^{(2)}_{g,(2,L)}(n_f+1) \Bigr]
     +~a_s^3 \Bigl[~A_{Qg}^{(3)}(n_f+1)~\delta_2
     +~A_{Qg}^{(2)}(n_f+1)~C^{(1), {\sf NS}}_{q,(2,L)}(n_f+1)
     \N\\ &&
     +~A_{gg,Q}^{(2)}(n_f+1)~\Ctil^{(1)}_{g,(2,L)}(n_f+1)
     +~A_{Qg}^{(1)}(n_f+1)\Bigl\{
     C^{(2), {\sf NS}}_{q,(2,L)}(n_f+1)
     \N\\ &&      +~\Ctil^{(2), {\sf PS}}_{q,(2,L)}(n_f+1)\Bigr\}
     +~A_{gg,Q}^{(1)}(n_f+1)~\Ctil^{(2)}_{g,(2,L)}(n_f+1)
     +~\Ctil^{(3)}_{g,(2,L)}(n_f+1) \Bigr]~, 
\N\\
\label{eqWIL5}
\end{eqnarray}
with $\delta_2 = 0 (1)$ for $F_L (F_2)$ and $\hat{f}(n_f) = f(n_f+1) - f(n_f), \tilde{f}(n_f) = 
f(n_f)/n_f$.

The renormalized massive OMEs depend on the ratio $m^2/\mu^2$, while the scale ratio in the massless 
Wilson coefficients is $\mu^2/Q^2$. The latter are pure functions of the momentum fraction $z$, or the
Mellin variable $N$, if one sets $\mu^2 = Q^2$. The mass dependence of the heavy flavor Wilson 
coefficients in the asymptotic region derives from the unrenormalized massive OMEs
\begin{eqnarray}
\label{eq:unep}
\Ahat^{(3)}_{ij}(\ep)  = 
  \frac{1}{\ep^3} \hat{a}^{(3),3}_{ij} 
  + \frac{1}{\ep^2} \hat{a}^{(3),2}_{ij} 
  + \frac{1}{\ep} \hat{a}^{(3),1}_{ij} 
  + \hat{a}^{(3),0}_{ij}~,
\end{eqnarray}
applying mass, coupling constant, and operator-renormalization, as well as mass factorization, 
cf.~Ref.~\cite{Bierenbaum:2009mv}. 

The renormalized massive OMEs obey then the general structure
\begin{eqnarray}
\label{eq:log}
A^{(3)}_{ij}\left(\frac{m^2}{Q^2}\right)  = 
  a^{(3),3}_{ij} \ln^3\left(\frac{m^2}{Q^2}\right)
+ a^{(3),2}_{ij} \ln^2\left(\frac{m^2}{Q^2}\right)
+ a^{(3),1}_{ij} \ln\left(\frac{m^2}{Q^2}\right)
+ a^{(3),0}_{ij}~.
\end{eqnarray}
The subsequent calculations will be performed in the $\overline{\rm MS}$ scheme. For other scheme choices 
see Ref.~\cite{Bierenbaum:2009mv}. Therefore the strong coupling constant is obtained as the {\it 
perturbative} solution of the equation
\begin{eqnarray}
\label{eq:as}
\frac{d a_s(\mu^2)}{d \ln(\mu^2)} = - \sum_{l=0}^\infty \beta_l a_s^{l+2}(\mu^2)
\end{eqnarray}
to 3--loop order, where $\beta_k$ are the expansion coefficients of the QCD $\beta$--function and $\mu^2$ 
denotes the renormalization scale. For simplicity we identify
the factorization and renormalization scales in the following.

\vspace*{-1mm}
\section{The Massive Operator Matrix Elements}
\label{sec:OME}

\noindent
The operator matrix elements $\propto n_f$ for both $F_2(x,Q^2)$ and transversity are obtained 
by the massive two-loop graphs inserting a further massless fermion line and new planar
three-loop topologies, cf.~\cite{Ablinger:2010ha,Klein:2009ig}, as well as 3-loop graphs containing 
bubble topologies with 
operator insertions linked either linked to massive or massless fermion lines, cf. 
\cite{Bierenbaum:2009mv}. The 
calculation  
was carried out in Feynman-gauge~\footnote{In Ref.~\cite{Bierenbaum:2009mv} we have 
kept the gauge parameter for part of the moments and found gauge independence. In the present 
calculation we have compared the results also on the basis of diagrams with the moments obtained there.}
using {\tt 
FORM}~\cite{FORM} and {\tt MAPLE}-codes, and 
applied the package {\tt color} \cite{vanRitbergen:1998pn} for the color algebra. As in earlier cases 
\cite{TOL2} we computed the
Feynman parameter-integrals directly, without applying the integrating-by-parts method~\cite{IBP}.
The corresponding integrals can be mapped onto sums over generalized hypergeometric functions prior the 
$\ep$--expansion, which allow to obtain the Laurent series in $\ep$. Finally, up to three-fold nested
sums over hypergeometric expressions, equipped with harmonic sums, have to be performed, for which the 
package {\tt Sigma}~\cite{SIGMA}, constructing difference and product fields, was applied and extended. 
Some details of the computation are presented in Appendix~A.

The massive OMEs $A_{ij}^{(k)}(N)$ are finally obtained as rational functions of the Mellin variable $N$, 
multiple zeta 
values~\cite{Blumlein:2009cf}, and nested harmonic sums~\cite{HSUM1,HSUM2}.
The latter are defined recursively by
\begin{eqnarray}
S_{b,\vec{a}}(N) = \sum_{k=1}^N \frac{({\rm sign}(b))^k}{k^{|b|}} 
S_{\vec{a}}(k),~~~~~S_{\emptyset}(N) = 1~.
\end{eqnarray}
As a short-hand notation we use $S_{\vec{a}}(N) \equiv S_{\vec{a}}$. In representing the results, the 
algebraic relations of the nested harmonic sums~\cite{Blumlein:2003gb} are applied. In the following we 
present the constant contributions to the unrenormalized OMEs (\ref{eq:unep}) as genuine quantities, to 
allow for different scheme choices, cf. Ref.~\cite{Bierenbaum:2009mv}.
\subsection{\boldmath Operator Matrix Elements contributing to $F_2(x,Q^2)$}
\label{sec:OME1}

\vspace*{1mm}
\noindent
The $O(n_f)$ contribution to the unrenormalized OME $\Ahat_{Qg}^{(3)}(\ep,N)$, $\hat{a}_{Qg}^{(3),0}$, 
reads~:
\begin{eqnarray}
\label{EQ:1}
 \hat{a}_{Qg}^{(3),0}&=& 
    n_fT_F^2C_A\Biggl\{
         \frac{16(N^2+N+2)}{27N(N+1)(N+2)}\Bigl[
             108 S_{-2,1,1}
            -78 S_{2,1,1}
            -90 S_{-3,1}
            +72 S_{2,-2}
            -6 S_{3,1}
\N\\ && \hspace{-12mm}
            -108 S_{-2,1} S_1
            +42 S_{2,1} S_1
            -6 S_{-4}
            +90 S_{-3} S_1
            +118 S_3 S_1
            +120 S_4
            +18 S_{-2} S_2
            +54 S_{-2} S_1^2
\N\\ && \hspace{-12mm}
            +33 S_2 S_1^2
            +15 S_2^2
            +2 S_1^4
            +18S_{-2}\zeta_2
            +9S_2\zeta_2
            +9S_1^2\zeta_2
            -42 S_1\zeta_3
                                    \Bigr]
\N\\ && \hspace{-12mm}
        +32\frac{5N^4+14N^3+53N^2+82N+20}{27N(N+1)^2(N+2)^2}\Bigl[
             6 S_{-2,1}
            -5 S_{-3}
            -6 S_{-2}S_1
                                    \Bigr]
\N\\ && 
\hspace{-12mm}
        -\frac{64(5N^4+11N^3+50N^2+85N+20)}{27N(N+1)^2(N+2)^2}S_{2,1}
\N\\ && \hspace{-12mm}
        -\frac{16(40N^4+151N^3+544N^2+779N+214)}{27N(N+1)^2(N+2)^2}S_2S_1
\N\\ && \hspace{-12mm}
        -\frac{32(65N^6+429N^5+1155N^4+725N^3+370N^2+496N+648)}{81(N-1)N^2(N+1)^2(N+2)^2}S_3
\N\\ && \hspace{-12mm}
        -\frac{16(20N^4+107N^3+344N^2+439N+134)}{81N(N+1)^2(N+2)^2}S_1^3
        +\frac{Q_{1}(N)}{81(N-1)N^3(N+1)^3(N+2)^3} S_2
\N\\ && \hspace{-12mm}
        +\frac{32(47N^6+278N^5+1257N^4+2552N^3+1794N^2+284N+448)}{81N(N+1)^3(N+2)^3}S_{-2}
\N
\end{eqnarray}\begin{eqnarray}
 && \hspace{-12mm}
        +\frac{8(22N^6+271N^5+2355N^4+6430N^3+6816N^2+3172N+1256)}{81N(N+1)^3(N+2)^3}S_1^2
\N\\
 && \hspace{-12mm}
        +\frac{Q_{2}(N)}{243(N-1)N^2(N+1)^4(N+2)^4} S_1
        +\frac{448(N^2+N+1)(N^2+N+2)}{9(N-1)N^2(N+1)^2(N+2)^2}\zeta_3
\N\\
&& \hspace{-12mm}
        -\frac{16(5N^4+20N^3+59N^2+76N+20)}{9N(N+1)^2(N+2)^2}S_1\zeta_2
        -\frac{Q_{3}(N)}{9(N-1)N^3(N+1)^3(N+2)^3} \zeta_2
\N\\
&& \hspace{-12mm}
        -\frac{Q_{4}(N)}{243(N-1)N^5(N+1)^5(N+2)^5}
       \Biggr\}
\N\\
&& \hspace*{-12mm}
   +n_fT_F^2C_F\Biggl\{
         \frac{16(N^2+N+2)}{27N(N+1)(N+2)}\Bigl[
             144S_{2,1,1}
            -72S_{3,1} 
            -72S_{2,1}S_1
            +48S_4
            -16S_3S_1
\N
\\ 
&& \hspace*{-12mm}
            -24S_2^2
            -12S_2S_1^2
            -2S_1^4
            -9 S_1^2\zeta_2
            +42S_1\zeta_3
                                    \Bigr]
        +32\frac{10N^3+49N^2+83N+24}{81N^2(N+1)(N+2)}\Bigl[
             3S_2S_1
             +S_1^3
                                    \Bigr]
\N\\ && \hspace{-12mm}
        -\frac{128(N^2-3N-2)}{3N^2(N+1)(N+2)}S_{2,1}
        -\frac{Q_{5}(N)}{81(N-1)N^3(N+1)^3(N+2)^2} S_3
\N\\ && \hspace{-12mm}
        +\frac{Q_{6}(N)}{27(N-1)N^4(N+1)^4(N+2)^3} S_2
        -\frac{32(10N^4+185N^3+789N^2+521N+141)}{81N^2(N+1)^2(N+2)}S_1^2
\N\\ && \hspace{-12mm}
        -\frac{16(230N^5-924N^4-5165N^3-7454N^2-10217N-2670)}{243N^2(N+1)^3(N+2)}S_1
\N\\ && \hspace{-12mm}
        +\frac{16(5N^3+11N^2+28N+12)}{9N^2(N+1)(N+2)}S_1\zeta_2
        -\frac{Q_{7}(N)}{9(N-1)N^3(N+1)^3(N+2)^2}\zeta_3
\N\\ && \hspace{-12mm}
        +\frac{Q_{8}(N)}{9(N-1)N^4(N+1)^4(N+2)^3}\zeta_2
        +\frac{Q_{9}(N)}{243(N-1)N^6(N+1)^6(N+2)^5} 
        \Biggr\}~,\label{aQg3}
\end{eqnarray}
with the polynomials
\begin{eqnarray} 
 Q_{1}(N)&=&32N^9-936N^8+6448N^7+55208N^6+126160N^5+61760N^4
\N\\
&&
-53152N^3-25024N^2-32256N -13824
~, \\
 Q_{2}(N)&=&+7856N^{10}+84672N^9+377648N^8+985568N^7+1395456N^6
\N\\
&&
+470688N^5-1183712N^4-1180224N^3-182528N^2-42752N
\N\\
&&
+13824
~,
\\
 Q_{3}(N)&=&60N^9+360N^8+584N^7-128N^6-2004N^5-2440N^4-976N^3
\N\\
&&
-192N^2+896N+384
          ~, \\
 Q_{4}(N)&=&28776N^{15}+356112N^{14}+1896088N^{13}+5538320N^{12}+9112264N^{11}
\N\\
&&
+6793968N^{10}-3019528N^9-11879520N^8-11673088N^7 
\N\\
&&
-6450992N^6-3726976N^5-2248128N^4-183296N^3+268032N^2
\N\\
&&
+147456N+27648
          ~, 
\\
 Q_{5}(N)&=&+464N^8-15616N^7-38112N^6+27776N^5+146064N^4+119552N^3
\N\\
&&
+109312N^2+86016N+62208~, 
\\
 Q_{6}(N)&=&456N^{11}+4376N^{10}+11328N^9-3184N^8-54552N^7-111720N^6
\N\\
&&
-155376N^5-251072N^4-312192N^3-222464N^2-135936N
\N\\
&&
-41472
          ~, \N\\
Q_{7}(N)&=&168N^8+672N^7+784N^6-3192N^4-5600N^3-7168N^2-4480N
\N
\end{eqnarray}\begin{eqnarray}
&&
-2688 ~,
\\
 Q_{8}(N)&=&90N^{11}+630N^{10}+1592N^9+1260N^8-1934N^7-8218N^6
\\
&&
-15524N^5-23944N^4-26752N^3-18400N^2-11328N-3456
          ~, \\
 Q_{9}(N)&=&15777N^{17}+186525N^{16}+879391N^{15}+1874085N^{14}+575913N^{13}
\N\\
&&
-5568833N^{12}-10465411N^{11}-2970289N^{10}+11884298N^9
\N\\
&&
+12640320N^8-10343664N^7-40750480N^6-55711424N^5
\N\\
&&
-53947712N^{4}-42534912N^3-23256576N^2-7865856N
\N\\
&&
-1244160
          ~,
\end{eqnarray}
and
\begin{eqnarray}
\zeta_k = \sum_{l=1}^\infty \frac{1}{l^k},~k~\in~{\mathbb N},~~k \geq 2
\end{eqnarray}
denotes the Riemann $\zeta$--function.

The corresponding contribution to the pure singlet OME $\Ahat_{Qq}^{\sf PS, (3)}(\ep,N)$ is given by
\begin{eqnarray} 
\label{EQ:2}
\hat{a}_{Qq}^{\sf PS, (3),0}&=& \frac{n_f T_F^2\,C_F} {N^{2} (1+N) ^{2} (2+N
)  (N-1) } \Biggl\{
(N^2+N+2)^2\Biggl(
-{\frac {1760}{27}}\,S_3
-{\frac {208}{9}}\,S_2 S_1
-{\frac {16}{27}}\, S_1^{3}
\N\\
&& 
-\frac {16}{3}\,S_1\zeta_2
+\frac {224}{9}\,\zeta_3
\Biggr)
+
{\frac {Q_{10}(N)}{N (1+N)(2+N) }} \left[
{\frac {208}{27}}\,
S_2
+{\frac {16}{27}}\,S_1^2
+{\frac {16}{9}}\,
\zeta_2 \right] \N\\ &&
-{\frac {32}{81}}\,{\frac { Q_{11}(N)}{N^{2} (1+N)^{2} (2+N)^{2}}} S_1
+{\frac {32}{243}}\,{\frac {Q_{12}(N)}{N^{3} (1+N) ^{3} (2+N) ^{3}}}
\Biggr\}~,\label{aQqPS3}
\end{eqnarray}
with
{\begin{eqnarray} 
Q_{10}(N)&=& 8\,N^{7}+37\,N^{6}+68\,N^{5}-11\,N^{4}-86\,N^{3}-56\,N^{2}-104\,N-48~,
\\
Q_{11}(N)&=& 25\,N^{10}+176\,N^{9}+417\,N^{8}+30\,N^{7}-20\,N^{6}+1848\,N^{5}
\N\\
&&
+2244\,N^{4}+1648\,N^{3}+3040 N^2+2112 N +576~,
\\
Q_{12}(N)&=& 158\,N^{13}+1663\,N^{12}+7714\,N^{11}+23003\,N^{10}+56186\,N^{9}
\N\\
&&
+89880\,N^{8}+59452\,N^{7}-8896\,N^{6}
-12856\,N^{5}-24944\,N^{4}
\N\\
&&
-84608\,N^{3}-77952\,N^{2}-35712\,N-6912~,
\end{eqnarray}

The second pure--singlet operator matrix element is $\Ahat_{qq,Q}^{\sf PS, (3)}(\ep,N)$. Its constant
term reads~:
\begin{eqnarray}
\label{EQ:3}
\hat{a}_{qq,Q}^{\sf PS, (3),0}&=&
\frac{n_f T_F^2\,C_F} {N^2 (N-1)  (2+N) (1+N)^2} \Biggl\{
(N^2+N+2)^2\Biggl(
{\frac {256}{27}}\,S_3
+{\frac {128}{9}}\,S_2 S_1
\N\\
&& 
+{\frac {128}{27}}\,S_1^3
+{\frac {32}{3}}\,S_1 \zeta_2
+{\frac {224}{9}}\,\zeta_3
\Biggr)
- {\frac {Q_{13}(N)}{N (2+N) (1+N)}} \left[
{\frac {64}{27}}\, S_2
+ {\frac {64}{27}}\, S_1^2
+ \frac{16}{9} \zeta_2 \right]
\N\\
&&
+{\frac {64}{81}}\,{\frac {Q_{14}(N)}{N^{2} (2+N) ^{2} (1+N)^{2}}} S_1
-{\frac {32}{243}}\,{\frac {Q_{15}(N)}{{N}^{3} (2+N)^{3} (1+N)^{3}}}\Biggr\}~,\label{aqqQPS3}
\end{eqnarray}
with
\begin{eqnarray}
\N\\
Q_{13}(N)&=& 16\,N^{7}+74\,N^{6}+181\,N^{5}+266\,N^{4}+269\,N^{3}+230\,N^{2}
\N\\
&&
+44\,N-24~, 
\end{eqnarray}\begin{eqnarray}
Q_{14}(N)&=&  181\,N^{10}+1352\,N^{9}+4737\,N^{8}+10101\,N^{7}
\N\\
&&
+14923\,N^{6}+17085\,N^{5}+14133\,N^{4}+5944\,N^{3}+568\,N^{2}-48\,N
+144~,\\
Q_{15}(N)&=& 2074\,N^{13}+21728\,N^{12}+105173\,N^{11}+311482\,N^{10}+636490\,N^{9}
\N\\
&&
+966828\,N^{8}+1126568\,N^{7}+968818\,N^{6}+550813\,N^{5}
\N\\
&&
+169250\,N^{4}+12104\,N^{3}-3408\,N^{2}-1008\,N-864~.
\end{eqnarray}

The constant term of the unrenormalized flavor non--singlet  operator matrix element  $\Ahat_{qq,Q}^{\sf 
NS,  (3)}(\ep,N)$ is given by~:
\begin{eqnarray} 
\label{EQ:4}
\hat{a}_{qq,Q}^{\sf NS, (3),0}&=&
n_f T_F^2\,C_F 
\Biggl\{
{\frac {64}{27}}\,\,S_4
+{\frac {448}{27}}\,\zeta_3 S_1
+{\frac {32}{9}}\,\zeta_2\,S_2
-{\frac {320}{81}}\,\,S_3
\N\\
&&
-{\frac {160}{27}}\,\zeta_2\,S_1
-{\frac {112}{27}}\,{\frac {3\,N^2+3\,N+2}{
    (1+N) N}} \zeta_3
+\frac {640}{27}\,\,S_2
\N\\
&&
+\frac {4}{27}\,\frac {
                      3\,N^{4}
                      +6\,N^{3}
                      +47\,N^{2}
                      +20\,N
                      -12}{(1+N)^2 N^{2}} \zeta_2
-\frac {55552}{729}\,S_1
\N\\
&&
+\frac {2}{729}\,{\frac {Q_{16}(N)}{(1+N) ^{4}N^{4}}}
\Biggr\}~,\label{aqqQNS3}
\end{eqnarray}
where
\begin{eqnarray}
Q_{16}(N)&=&
11751\,N^{8}+47004\,N^{7}+93754\,N^{6}+104364\,N^{5}+55287\,N^{4}
\N\\
&&
+6256\,N^{3}-2448\,N^{2}-144\,N-432~.
\end{eqnarray}

Finally, the constant contribution to $\Ahat_{qg,Q}^{(3)}(\ep,N)$ at $O(n_f)$ is~: 
\begin{eqnarray}
\label{EQ:5}
\hat{a}_{qg,Q}^{(3),0}&=&
\frac {n_f T_F^2} {N (N+1) (N+2)}
\N\\
&&
\Biggl\{
C_F 
\Bigl[
(N^2+N+2)
\Bigl(
\frac{4}{27}  S_1^4
+\frac{8}{3}  \zeta _2 S_1^2
+\frac{8}{9}  S_2 S_1^2
+\frac{224}{9}  \zeta_3 S_1
+\frac{32}{27}  S_3 S_1
+\frac{4}{9}  S_2^2
\N
\\
&&
+8 \zeta _2 S_2 
+\frac{40}{9}  S_4 
-\frac{56 Q_{17}(N)}{9 (N-1) N^2 (N+1)^2 (N+2)} \zeta_3 
\Bigr)
\N
\\
&&
-\frac{16 (10 N^3+13 N^2+29 N+6) S_1^3}{81 N}
+\frac{8 (215 N^4+481 N^3+930 N^2+748 N+120)}{81 N (N+1)} S_1^2
\N\\
&&
-\frac {16 (10 N^3+13 N^2+29 N+6)} {27 N}
\Bigl(
3 \zeta_2 S_1
+ S_2 S_1
\Bigr)
-\frac{32 (40 N^3+61 N^2+89 N+6) }{81 N} S_3
\N\\
&&
+\frac{8 (221 N^4+515 N^3+814 N^2+548 N+40)}{27 N (N+1)} S_2
+\frac{4 Q_{18}(N) }{9 (N-1) N^3 (N+1)^3 (N+2)^2} \zeta _2
\N\\
&&
-\frac{16 Q_{19}(N)}{243 N (N+1)^2} S_1
+\frac{8 Q_{20}(N)}{243 (N-1) N^5 (N+1)^5 (N+2)^4}
\Bigr]
\N\\
&&
+ C_A \Bigl[
(N^2+N+2) \Bigl(
-\frac{4}{27}  S_1^4
-\frac{8}{3}  \zeta _2 S_1^2
+\frac{8}{9}  S_2 S_1^2
-\frac{56}{9}  S_4
-\frac{128}{9}  S_{3,1}
+\frac{64}{9}  S_{2,1,1}
\N
\\
&&
+\frac{160}{27}  S_3 S_1
-\frac{64}{9}  S_{2,1} S_1
-\frac{4}{9}  S_2^2
-\frac{128}{9}  S_{-4}
-\frac{224}{9}  \zeta_3 S_1
-\frac{16}{3}  \zeta_2 S_{-2}
-\frac{8}{3}  \zeta_2 S_2
\N\\
&&
+\frac{448 (N^2+N+1)}{9 (N-1) N (N+1) (N+2)} \zeta_3
\Bigr)
+
\frac{32 (5 N^4+20 N^3+41 N^2+49 N+20) }{81 (N+1) (N+2)} \times 
\N
\end{eqnarray}\begin{eqnarray}
&&
\times \Bigl(
9 \zeta_2 S_1
-3 S_2 S_1
+ 12 S_{2,1}
+ S_1^3
\Bigr)
+\frac{64 \left(5 N^4+38 N^3+59 N^2+31 N+20\right)}{81 (N+1) (N+2)} S_3
\N\\
&&
+\frac{128}{27} (5 N^2+8 N+10) S_{-3}
-\frac{8 Q_{21}(N)}{81 (N+1)^2 (N+2)^2} S_1^2
\N\\
&&
+\frac{8 Q_{22}(N)}{9 (N-1) N^2 (N+1)^2 (N+2)^2} \zeta_2
-\frac{32 (121 N^3+293 N^2+414 N+224)}{81 (N+1)} S_{-2}
\N\\
&&
-\frac{8 Q_{23}(N)}{81 (N+1)^2 (N+2)^2} S_2
+\frac{16 Q_{24}(N)}{243 (N-1) N (N+1)^3 (N+2)^3} S_1
\N\\
&&
+\frac{16 Q_{25}(N)}{243 (N-1) N^4 (N+1)^4 (N+2)^4}
\Bigr]
\Biggr\}~,
\end{eqnarray}
with
\begin{eqnarray}
Q_{17}(N)&=&3 N^6+9 N^5-N^4-17 N^3-38 N^2-28 N-24~,
\\
Q_{18}(N)&=&18 N^{11}+126 N^{10}+365 N^9+630 N^8+652 N^7+626 N^6+1309 N^5
\N\\
&&
+3170 N^4+4736 N^3+3584 N^2+2352 N+864~,
\\
Q_{19}(N)&=&2507 N^5+8076N^4+16120 N^3+18997 N^2+9898 N+1344~,
\\
Q_{20}(N)&=&2322 N^{17}+30186 N^{16}+177047 N^{15}+627060 N^{14}+1509207 N^{13}
\N\\
&&
+2623160 N^{12}+3436402 N^{11}+3728602 N^{10}+4151281 N^9
\N\\
&&
+5013306 N^8+5011065 N^7+3770902
   N^6+3291500 N^5+3951272 N^4
\N\\
&&
+3797616 N^3+2319264 N^2+862272 N+155520~,
\\
Q_{21}(N)&=&206 N^6+1361 N^5+4134 N^4+7577 N^3+8394 N^2+4868 N+1144~,
\\
Q_{22}(N)&=&6 N^9+36 N^8+11 N^7-257 N^6-825 N^5-1375 N^4-1396 N^3-984 N^2
\N\\
&&
-352 N-48~,
\\
Q_{23}(N)&=&332 N^6+2537 N^5+7848 N^4+13145 N^3+13122 N^2+7412 N+1720~,
\\
Q_{24}(N)&=&2228 N^{10}+19197 N^9+72518 N^8+155774 N^7+193362 N^6+94317 N^5
\N\\
&&
-87644 N^4-163656 N^3-91040 N^2-11888 N+3456~,
\\
Q_{25}(N)&=&2040 N^{15}+24480 N^{14}+116165 N^{13}+254533 N^{12}+78119 N^{11}
\N\\
&&
-1089300 N^{10}-3414794 N^9-5743128 N^8-6358562 N^7-4824553 N^6
\N\\
&&
-2448740 N^5-783540 N^4-213184 N^3-155568 N^2-97344 N
\N\\
&&
-22464~.
\end{eqnarray}
We compared 
$\hat{a}_{Qg}^{(3),0}(N)$, 
$\hat{a}_{Qq}^{\sf PS, (3),0}(N)$, 
$\hat{a}_{qq,Q}^{\sf PS, (3),0}(N)$, 
$\hat{a}_{qq,Q}^{\sf NS, (3),0}(N)$, and 
$\hat{a}_{qg,Q}^{(3),0}(N)$, Eqs.~(
\ref{EQ:1},
\ref{EQ:2},
\ref{EQ:3},
\ref{EQ:4},
\ref{EQ:5}), to the fixed moments computed in Ref.~\cite{Bierenbaum:2009mv}
and found agreement.

The OMEs $A_{qq,Q}^{\sf PS, (3)}(N)$ and $A_{qg,Q}^{\sf (3)}(N)$ receive contributions 
$\propto n_f T_F^2 C_{A,F}$ only. 
Due to this we present as well the constant parts of the renormalized OMEs. They read~:
\begin{eqnarray}
a_{qq,Q}^{\sf PS, (3),0} &=& n_f T_F^2 C_F \Biggl\{
\frac{(N^2+N+2)^2}{(N-1) N^2 (1+N)^2 (2+N)} \left[\frac{80}{27}\left(  S_1^3
                                                   + 3 S_1 S_2
                                                   + 2  S_3\right)  
                                                   +\frac{256}{9}  \zeta_3 \right]
\nonumber\\ & & 
 - \frac{16 R_1(N)}{27 (N-1) N^3 (1+N)^3 (2+N)^2} \left[S_1^2 +S_2\right] 
  + \frac{32 R_2(N)}{81 (N-1) N^4 (1+N)^4 (2+N)^3} S_1
\nonumber\\ & & 
  + \frac{R_3(N)}{243 (N-1) N^5 (1+N)^5 (2+N)^4} \Biggr\}
\label{eq:aqqQren}
\end{eqnarray}
with
\begin{eqnarray}
R_1(N) &=&
  40 N^7
+185 N^6
+430 N^5
+521 N^4
+452 N^3
+404 N^2
-16 N
-96~,    \\
R_2(N) &=& 
233 N^{10}
+1744 N^9
+5937 N^8
+11454 N^7
+14606 N^6
+15396 N^5
+12030 N^4
\nonumber
\\ &&
+3272 N^3
-928 N^2
-96 N
+288~, \\
R_3(N) &=& 
-42560 N^{13}
-445792 N^{12}
-2124448 N^{11}
-6005792 N^{10}
-11345024 N^9
\nonumber\\ &&
-15758592 N^8
-17045248 N^7
-13567040 N^6
-6545312 N^5
-1096768 N^4
\nonumber\\ &&
+374528 N^3
+109056 N^2
+32256 N
+27648~, 
\end{eqnarray}
and
\begin{eqnarray}
a_{qg,Q}^{\sf (3),0} &=&  n_f T_F^2 \Biggl\{C_F \Biggl[
     \frac{N^2+N+2}{N(N+1)(N+2)} \left[ -\frac{56}{9}  S_4 
                                        +\frac{32}{27} S_3 S_1
                                        +\frac{8}{9}   S_2 S_1^2
                                        +\frac{4}{9}   S_2^2
                                        +\frac{4}{27}  S_1^4 
                                        +\frac{256}{9} S_1 \zeta_3  \right]       
\N\\ &&
     -\frac{16 (10 N^3 + 13 N^2 + 29 N +6)}{81 N^2 (1+N) (2+N)} \left[S_1^3 + 3 S_2 S_1\right]
     +\frac{32 (5 N^3-16 N^2+N-6)}{81 N^2 (1+N) (2+N)} S_3
\N\\ &&
+ \frac{8 (109 N^4 +291 N^3 +478 N^2 +324 N + 40)} {27 N^2 (1+N)^2 (2+N)} S_2
\N\\ &&
+ \frac{8 (215 N^4 +481 N^3 +930 N^2 +748 N + 120)}{81 N^2 (1+N)^2 (2+N)} S_1^2
- \frac{R_4(N)}{243 N^2 (1+N)^3 (2+N)} S_1 
\N\\ &&
-\frac{64 (N^2+N+2) R_5(N)}{9 (N-1) N^3 (1+N)^3 (2+N)^2} \zeta_3
+ \frac{R_6(N)}{243 (N-1) N^6 (1+N)^6 (2+N)^5} \Biggr]
\N\\ &&
+ C_A \Biggl[
\frac{N^2+N+2}{N (N+1) (N+2)} \left[ 
- \frac{56}{9} S_4 
- \frac{128}{9} S_{-4} 
+ \frac{160}{27} S_3 S_1 
- \frac{4}{9} S_2^2  
+ \frac{8}{9} S_2 S_1^2 \right. 
\nonumber\\ && \left.
- \frac{4}{27} S_1^4  
- \frac{64}{9} S_{2,1} S_1   
- \frac{128}{9} S_{3,1}   
+ \frac{64}{9} S_{2,1,1}   
- \frac{256}{9} \zeta_3 S_1 \right]   
\N\\ &&
+\frac{32 (
5 N^4
+20 N^3
+41 N^2
+49 N
+20
)}{81 N (1+N)^2 (2+N)^2}
 \left[S_1^3
      + 12 S_{2,1} 
      -  3 S_2 S_1 \right]
\N\\ &&
+\frac{64}{81} \frac{(
  5 N^4
+38 N^3
+59 N^2
+31 N
+20
)}{N (1+N)^2 (2+N)^2} S_3
+\frac{128}{27} \frac{(5 N^2 + 8 N +10)}{N (1+N) (2+N)} S_{-3}
\N\\ &&
+\frac{512}{9} \frac{(N^2+N+1) (N^2+N+2)}{(N-1) N^2 (1+N)^2 (2+N)^2} \zeta_3
- \frac{16 R_7(N)}{81 N (1+N)^3 (2+N)^3} S_2
\N\\ &&
- \frac{32 (
 121 N^3
+293 N^2
+414 N
+224
)}{81 N (1+N)^2 (2+N)} S_{-2}
- \frac{R_8(N)}{81 N (1+N)^3 (2+N)^3} S_1^2
\N\\ &&
+\frac{16 R_9(N)}{243 (N-1) N^2 (1+N)^4 (2+N)^4} S_1
+ \frac{8 R_{10}(N)}{243 (N-1) N^5 (1+N)^5 (2+N)^5} \Biggr]\Biggr\}~,
\label{eq:aqgQren}
\end{eqnarray}
with
\begin{eqnarray}
R_4(N) &=& 
 24368 N^5
+81984 N^4
+179200 N^3
+225232 N^2
+126880 N 
+21504~, 
\\
R_5(N) &=&
 3 N^6
+9 N^5
-N^4
-17 N^3
-38 N^2
-28 N
-24~, \\
R_6(N) &=&
 13923 N^{17}
+180999 N^{16}
+1064857 N^{15}
+3812487 N^{14}
+9348807 N^{13}
\nonumber\\ &&
+16391845 N^{12}
+20248499 N^{11}
+17070917 N^{10}
+11536274 N^9
+11303496 N^8
\nonumber\\
&&
+13846104 N^7
+16104128 N^6
+22643488 N^5
+29337472 N^4
\nonumber
\end{eqnarray}
\begin{eqnarray}
&&
+26395008 N^3
+15388416 N^2
+5612544 N
+995328~,
\\
R_7(N) &=&
  139 N^6
+1093 N^5
+3438 N^4
+5776 N^3
+5724 N^2
+3220 N
+752~,
\\
R_8(N) &=&
  1648 N^6
+11104 N^5
+34368 N^4
+63856 N^3
+71904 N^2
+43264 N
+10880~,
\\
R_9(N) &=&
+1244 N^{10}
+10557 N^9
+40547 N^8
+90323 N^7
+114495 N^6
+49344 N^5
\nonumber\\ &&
-69902 N^4
-115200 N^3
-64352 N^2
-11264 N
+864~,
\\
R_{10}(N) &=&
3315 N^{15}
+39780 N^{14}
+194011 N^{13}
+471164 N^{12}
+416251 N^{11} 
-860568 N^{10}
\nonumber\\ &&
-3525799 N^9
-6015120 N^8
-6333994 N^7
-4373672 N^6
-1907512 N^5
\nonumber\\ &&
-499824 N^4
-217952 N^3
-264192 N^2
-160128 N
-34560~.
\end{eqnarray}
In both the constant terms of the renormalized OMEs Eq.~(\ref{eq:aqqQren},\ref{eq:aqgQren}) 
$\zeta_2$ does not contribute anymore.
Phenomenological applications of the corresponding massive Wilson coefficients 
are given in Ref.~\cite{BBKW}. 
\subsection{\boldmath The Operator Matrix Elements for Transversity}
\label{sec:OME2}

\vspace*{1mm}
\noindent
Transversity is a twist-2 flavor non--singlet operator matrix element related to a tensor operator, which 
cannot be accessed in 
deep-inelastic scattering, but via polarized semi-inclusive deep-inelastic scattering and the polarized 
Drell-Yan process. The anomalous dimensions for transversity are known to NLO~\cite{ANDIM3} and for
a series of moments to 3--loop order~\cite{Gracey:2003yrxGracey:2003mrxGracey:2006zrxGracey:2006ah}.
Phenomenological aspects of transversity have been reviewed in Ref.~\cite{RATCL}. The moments $N= 1 ... 13$
of the 3--loop massive OME were calculated in \cite{Blumlein:2009rg}. Similar to the flavor non--singlet
massive OME in the vector case we computed the $O(n_f)$ contributions for the transversity operator.
The constant part of the unrenormalized 3--loop OME is given by
\begin{eqnarray}
\hat{a}_{qq,Q}^{\sf TR, (3),0} &=&
n_f T_F^2 C_F \Biggl\{
\frac{64 }{27} S_4
+\frac{448 }{27} \zeta_3 S_1
+\frac{32 }{9} \zeta_2 S_2
-\frac{320 }{81} S_3
-\frac{160 }{27} \zeta_2 S_1
\N\\
&&
-\frac{112} {9} \zeta_3
+\frac{640}{27} S_2
+\frac {4} {9} \zeta_2
-\frac{55552}{729} S_1
\N\\
&&
+\frac {2 (3917 N^4 + 7834 N^3 +4157 N^2 -48 N -144)} {243 N^2 (1+N)^2}
\Biggr\}\label{aTrans}~.
\end{eqnarray}
The expression for general values of $N$ agrees with the corresponding contributions to the moments
calculated in \cite{Blumlein:2009rg} before. It is interesting to 
note that for this color factor the vector and tensor operators (\ref{aqqQNS3},\ref{aTrans}) lead to the 
same structures in the harmonic sums as for $\hat{a}_{qq,Q}^{\sf NS,(3),0}$.
\subsection{The Mathematical Structure of the Operator Matrix Elements}
\label{sec:HS1}

\vspace*{2mm}
\noindent
The $n_f T_F^2 C_{F,A}$--contributions at $O(a_s^3)$ to the massive operator matrix elements
contain nested harmonic sums up to weight {\sf w = 4}. This also applies to all individual
Feynman diagrams,~cf.~\cite{WISSB}. In intermediary results, generalizations of harmonic sums occur, 
see Appendix~\ref{app1}. As has been observed in the computation of various
other physical quantities before, such as anomalous dimensions and massless Wilson coefficients
to 3-loop order \cite{ANDIM2b,Vermaseren:2005qc,Blumlein:2009tj}, 
unpolarized and polarized massive OMEs to 2--loop order \cite{TOL2}, 
the polarized and unpolarized Drell-Yan  and Higgs-boson production cross section, time-like Wilson 
coefficients, and virtual- and soft corrections to Bhabha-scattering \cite{BRK}, the classes of contributing 
harmonic sums are always the same. They depend on the loop-order and the topologies of Feynman diagrams involved. 

In the present case the following harmonic sums emerge~:
\begin{eqnarray}
\label{eq:HSUM1}
& & S_1  \nonumber \\
& & S_2,~S_{-2}  \nonumber \\
& & S_3,~S_{-3},~S_{2,1},~S_{-2,1}  \nonumber \\
& & S_4,~S_{-4},~S_{3,1},~S_{-3,1},~S_{-2,2},~S_{2,1,1},~S_{-2,1,1}~.  
\end{eqnarray}
Note that this class, as for the other processes mentioned above, does not contain the index $\{-1\}$.
Moreover, we used the algebraic relations between the harmonic sums, cf.~\cite{Blumlein:2003gb}. Furthermore,
structural relations exist between harmonic sums, cf.~\cite{Blumlein:2009ta,STRUC_H}, 
which reduce the set (\ref{eq:HSUM1}) further. Here  the sums
\begin{eqnarray}
\label{eq:HSUM2}
S_{-2,2},~~S_{3,1} 
\end{eqnarray}
are connected by differential relations w.r.t.\ their argument $N$ to other sums of (\ref{eq:HSUM1}).
This is also the case for all single harmonic sums $S_{\pm n},~~~n \in \mathbb{N}$, $n > 1,$ using 
both 
the differentiation and argument-duplication relation, cf. \cite{HSUM1}. Due to this $S_1$ represents the
class of all single harmonic sums. I.e. only the {\it six} basic harmonic sums 
\begin{eqnarray}
\label{eq:HSUM3}
& & S_1  \nonumber \\
& & S_{2,1},~S_{-2,1}  \nonumber \\
& & S_{-3,1},~S_{2,1,1},~S_{-2,1,1}  
\end{eqnarray}
are needed to represent the 3-loop results for the 
$n_f T_F^2 C_{F,A}$--contributions to the OMEs calculated in the present paper. In the final 
representation we refer to the algebraic basis (\ref{eq:HSUM1}) and consider the basis (\ref{eq:HSUM3})
for a later numerical implementation. We sorted the respective expressions keeping a rational function
in $N$ in front of the harmonic sums (\ref{eq:HSUM1}) and $\zeta$--values, like $\zeta_{2}$ and $\zeta_3$.

The harmonic sums emerge from the series--expansion of hypergeometric structures like the
Euler $B$-- and $\Gamma$--functions and the Pochhammer--symbols in the (generalized) hypergeometric
functions $_PF_Q(a_i(\ep), b_i(\ep);1)$ in the dimensional parameter $\varepsilon$. This leads to single 
harmonic
sums first, which, through summation, turn into (multiple) zeta values \cite{Blumlein:2009cf} and nested
harmonic sums \cite{HSUM1,HSUM2}. The principle steps on the way from single--scale
Feynman diagrams to these structures have been described in Ref.~\cite{Blumlein:2009ta}. 

For phenomenological applications the heavy flavor corrections to the structure functions have to be 
known in $x$--space. Both the evolution of the parton densities and the Wilson coefficients have to be 
computed at complex values of $N$. The Mellin--inversion is then performed by a numerical contour 
integral around the singularities of the problem \cite{DFLM}. The analytic continuation
of the harmonic sums to complex values of $N$ is outlined in Refs.~\cite{ANCONT,Blumlein:2009ta,STRUC_H}. 
\subsection{\boldmath The OMEs in the Small and Large $x$ Region}
\label{sec:limit}

\vspace{2mm}
\noindent
In the small $x$ limit the following leading behaviour of the $\hat{a}_{ij}^{\sf (3),0}$,
${a}_{qq,Q}^{\sf PS, (3),0}$ and ${a}_{qg,Q}^{\sf (3),0}$ is obtained~:
\begin{eqnarray}
\hat{a}_{Qg}^{\sf (3),0}    &\propto& n_f T_F^2 \left\{ C_A 
\left[- \frac{18400}{729} + \frac{448}{27} \zeta_3 + \frac{16}{9} \zeta_2                               
\right]
+C_F \left[ - \frac{185408}{729} + \frac{896}{27} \zeta_3 - \frac{736}{27} \zeta_2                               
\right] \right\}\frac{1}{x} \nonumber\\
\\
\hat{a}_{Qq}^{\sf PS, (3),0}    &\propto& - n_f T_F^2 C_F \left[ - \frac{111104}{729}
                                                            + \frac{896}{27} \zeta_3
                                                              - \frac{320}{27} \zeta_2
                                                            \right] \frac{1}{x}
\end{eqnarray}\begin{eqnarray}
\hat{a}_{qq,Q}^{\sf PS, (3),0}  &\propto& - n_f T_F^2 C_F 
\left[ - \frac{111104}{729} + \frac{896}{27} \zeta_3 - \frac{320}{27} \zeta_2
\right]
\frac{1}{x}\\
{a}_{qq,Q}^{\sf PS, (3),0}  &\propto&  n_f T_F^2 C_F \frac{1024}{27} \left[ - \frac{47}{27} + \zeta_3 
\right] \frac{1}{x}\\
\hat{a}_{qq,Q}^{\sf NS, (3),0}  &\propto& - n_f T_F^2 C_F \frac{16}{81} 
\ln^3\left(\frac{1}{x}\right)  \\
\hat{a}_{qg,Q}^{\sf (3),0}    &\propto& n_f T_F^2 \left\{ C_A 
\left[- \frac{145408}{729} + \frac{448}{27} \zeta_3 - \frac{64}{3} \zeta_2                               
\right]
+C_F \left[ \frac{68608}{729} + \frac{896}{27} \zeta_3 + \frac{512}{27} \zeta_2                               
\right] \right\}\frac{1}{x} \nonumber\\
\\
{a}_{qg,Q}^{\sf (3),0}  &\propto& n_f T_F^2 \left\{
C_A \left[ - \frac{69472}{729}
+ \frac{512}{27} \zeta_3 \right]
+ C_F \left[ \frac{42688}{729} + \frac{1024}{27} \zeta_3 \right] \right\} \frac{1}{x}
\\
\hat{a}_{qq,Q}^{\sf TR, (3),0}  &\propto&  - n_f T_F^2 C_F \frac{32}{27} 
\ln\left(\frac{1}{x}\right)~. 
\end{eqnarray}
In case of the singlet and pure--singlet terms the leading behaviour is $\propto 1/x$, while
in the non--singlet cases it is logarithmic. The small-$x$ asymptotics
of $\hat{a}_{Qq}^{\sf PS, (3),0}$ and $\hat{a}_{qq,Q}^{\sf PS, (3),0}$ turn out to be the same. The 
matrix 
elements
are less singular than the leading terms in the Wilson coefficients,~cf.~\cite{SX,Vermaseren:2005qc}.

In the large $x$ limit one obtains the following leading behaviour, 
\begin{eqnarray}
\hat{a}_{Qg}^{\sf (3),0}       &\propto&   n_f T_F^2 (C_A - C_F) \frac{32}{27} \ln^4(1-x)\\ 
\hat{a}_{qq,Q}^{\sf NS, (3),0}  &\propto&  n_f T_F^2 C_F \left[ \frac{55552}{729} 
- 
\frac{448}{27} \zeta_3
+ \frac{160}{27} \zeta_2\right] \frac{1}{(1-x)_+} \\
\hat{a}_{qg,Q}^{\sf (3),0}, {a}_{qg,Q}^{\sf (3),0}     &\propto& - n_f T_F^2 (C_A - C_F) \frac{4}{27} 
\ln^4(1-x)  \\
\hat{a}_{qq,Q}^{\sf TR, (3),0}  &\propto&  n_f T_F^2 C_F \left[ \frac{55552}{729} - \frac{448}{27} 
\zeta_3
+ \frac{160}{27} \zeta_2\right] \frac{1}{(1-x)_+}~,
\end{eqnarray}
where
\begin{eqnarray}
\int_0^1 dx \frac{1}{(1-x)_+} f(x) = \int_0^1 dx \frac{f(x)-f(1)}{1-x}~,
\end{eqnarray}
and
\begin{eqnarray}
\hat{a}_{Qq}^{\sf PS, (3),0}      &\propto&  - n_f T_F^2 C_F \frac{16}{27} \frac{S_1^3}{N^2} \simeq
\frac{32}{9} \left[S_{1,3}(x) - \zeta_4\right] \\ 
\hat{a}_{qq,Q}^{\sf PS, (3),0}      &\propto&  n_f T_F^2 C_F \frac{128}{27} \frac{S_1^3}{N^2} \simeq
- \frac{256}{9} \left[S_{1,3}(x) - \zeta_4\right] \\
{a}_{qq,Q}^{\sf PS, (3),0}      &\propto&   n_f T_F^2 C_F \frac{80}{27} \frac{S_1^3}{N^2} \simeq
- \frac{160}{9} \left[S_{1,3}(x) - \zeta_4\right]~,
\end{eqnarray}
cf.~\cite{HSUM1,Blumlein:2004bs}.
In the latter case regular values are obtained
for $x \rightarrow 1$, where $S_{1,3}(x)$ denotes a Nielsen integral~\cite{NIELS},
\begin{eqnarray}
S_{n,p}(x) = (-1)^{n+p-1} \frac{1}{(n-1)! p!} \int_0^1 \frac{dz}{z} \ln^{(n-1)}(z) \ln^p(1-xz)~.
\end{eqnarray}
The large $x$ limits for $\hat{a}_{Qg}^{(3),0}$ and $\hat{a}_{qg,Q}^{(3),0}$, resp. $\hat{a}_{Qq}^{\sf 
PS, (3),0}$ 
and $\hat{a}_{qq,Q}^{\sf PS, (3),0}$ in the $n_f$ term differ by a factor of $-8$ and $-1/8$, while
the contributions to $\hat{a}_{qq,Q}^{\sf NS, (3),0}$ and $\hat{a}_{qq,Q}^{\sf TR, (3),0}$ are the same.
All terms are less singular compared to the massless cases \cite{Vermaseren:2005qc}.
\section{The Contributions to the Anomalous Dimensions}
\label{sec:andim}

\vspace{2mm}
\noindent
The anomalous dimensions appear in the $1/\ep$ term of the unrenormalized OMEs, see 
Ref.~\cite{Bierenbaum:2009mv}. As all other contributions to this term are known, they can 
be derived by comparing with the $1/\ep$~terms of the present computation. 
\subsection{\boldmath Vector Operators}
\label{sec:gam1}

\vspace*{1mm}
\noindent
From the OMEs
$\Ahat_{Qg}^{(3)}(\ep, N)$ and $\Ahat_{qg,Q}^{(3)}(\ep, N)$ one obtains~:
\begin{eqnarray} 
{\gamma}_{qg}^{\sf (2)}&=&\frac {n_f^2 T_F^2} {(N+1) (N+2)} \Biggl\{ 
C_A \Biggl[
\left(N^2+N+2\right) \Biggl(
\frac{128}{3 N} S_{2,1}
+\frac{32}{9 N} S_1^3
+\frac{128}{3 N} S_{-3}
+\frac{64}{9 N} S_3
\N\\
&&
-\frac{32 }{3 N} S_2 S_1
\Biggr)
-\frac{128 (5 N^2+8 N+10) }{9 N} S_{-2}
-\frac{64 (5 N^4+26 N^3+47 N^2+43 N+20) }{9 N (N+1) (N+2)} S_2
\N\\
&&
-\frac{64 (5 N^4+20 N^3+41 N^2+49 N+20) }{9 N (N+1) (N+2)} S_1^2
+\frac{64 P_{1}(N)}{27 N (N+1)^2 (N+2)^2} S_1
\N\\
&&
+\frac{16 P_{2}(N)}{27 (N-1) N^4 (N+1)^3 (N+2)^3}
\Biggr]
\N\\ &&
+ C_F \Biggl[ \frac{32}{9} \frac{N^2+N+2}{N} \left\{10 S_3 - S_1^3 - 3 S_1 S_2 \right\}
\N\\
&&
+\frac{32 (5 N^2+3 N+2) }{3 N^2} S_2
+\frac{32 (10 N^3+13 N^2+29 N+6) }{9 N^2} S_1^2
\N\\
&&
-\frac{32 (47 N^4+145 N^3+426 N^2+412 N+120)}{27 N^2 (N+1)} S_1
+\frac{4 P_{3}(N)}{27 (N-1) N^5 (N+1)^4 (N+2)^3}
\Biggr]
\Biggr\}\label{gammaqg}~,
\N\\
\end{eqnarray}
with
\begin{eqnarray} 
P_{1}(N)&=&19 N^6+124 N^5+492 N^4+1153 N^3+1362 N^2+712 N+152~,
\\
P_{2}(N)&=&165 N^{12}+1485 N^{11}+5194 N^{10}+8534 N^9+3557 N^8-8899 N^7
\N\\
&&
-10364 N^6+6800 N^5+25896 N^4+30864 N^3+19904 N^2
\N\\
&&
+7296 N+1152~,
\\
P_{3}(N)&=&99 N^{14}+990 N^{13}+4925 N^{12}+17916 N^{11}+46649 N^{10}+72446 N^9
\N\\
&&
+32283 N^8-95592 N^7-267524 N^6-479472 N^5-586928 N^4
\N\\
&&
-455168 N^3-269760 N^2-122112 N-27648~.
\end{eqnarray}
The $n_f^2$--contribution to the pure--singlet anomalous dimension results from
$\Ahat_{Qq}^{\sf PS, (3)}(\ep, N)$ and $\Ahat_{qq,Q}^{\sf PS, (3)}(\ep, N)$~: 
\begin{eqnarray} 
{\gamma}_{qq}^{\sf PS, (2)} &=& \frac {n_f^2 T_F^2 C_F} {(N-1) N^2 (N+1)^2 (N+2)} \Biggl\{
-{\frac {32}{3}}\,( {N}^{2}+N+2 )^{2} (S_1^{2}+S_2)
\N\\
&&
+{\frac {64}{9}}\,{\frac { P_{4}(N)\,}{{N} ( 1+N ) (2+N )}} S_1
-{\frac {64}{27}}\,{\frac {P_{5}(N)}{{N}^{2} ( 1+N ) ^{2} ( 2+N)^{2}}}
\Biggr\}\label{gammaqqPS}~,
\end{eqnarray}\begin{eqnarray}
P_{4}(N)&=&68\,{N}^{5}+37\,{N}^{6}+8\,{N}^{7}-11\,{N}^{4}-86\,{N}^{3}-56\,{N}^{2}-104\,N-48~, 
\\
P_{5}(N)&=&  
+52\,{N}^{10}
+392\,{N}^{9}
+1200\,{N}^{8}
+1353\,{N}^{7}
-317\,{N}^{6}
-1689\,{N}^{5}
\N\\
&&
-2103\,{N}^{4}
-2672\,{N}^{3}
-1496\,{N}^{2}
-48\,N
+144~.
\end{eqnarray}  
Both the $O(n_f)$ contributions to $\gamma_{qg}^{\sf (2)}$ and $\gamma_{qq}^{\sf PS, (2)}$ have thus 
been obtained by two independent new calculations.

The $n_f^2$--contribution in the flavor non--singlet case is derived from $\Ahat_{qq,Q}^{\sf NS, 
(3)}(\ep, N)$~:
\begin{eqnarray} 
{\gamma}_{qq}^{\sf NS, (2)} &=& n_f^2 T_F^2 C_F \left\{
{\frac {128}{9}}\,S_3
-{\frac {640}{27}}\,S_2
-{\frac {128}{27}}\,S_1
+{\frac {8}{27}}\,{\frac {P_{6}(N)}{{N}^{3} (1+N)^{3}}}\right\}\label{gammaqqNS}~,
\end{eqnarray}
with
\begin{eqnarray}
P_{6}(N)&=& 51\,{N}^{6}+153\,{N}^{5}+57\,{N}^{4}+35\,{N}^{3}+96\,{N}^{2}+16\,N-24 ~.
\end{eqnarray}  
The anomalous dimensions 
agree with the moments, resp. the general results, in 
Refs.~\cite{ANDIM1,ANDIM2b,Bierenbaum:2009mv}. Due to the algebraic 
compactification we obtain a lower number of harmonic sums $S_{\vec{a}}(N)$ if compared to 
Ref.~\cite{ANDIM2b}, and agree with \cite{Blumlein:2009tj}. For the flavor non--singlet case
the anomalous dimension has been predicted in \cite{ANDIM2a}.
\subsection{\boldmath Tensor Operator}
\label{sec:gam2}

\vspace*{1mm}
\noindent
The contribution to the transversity anomalous dimension $\propto n_f^2$ is obtained from
the single pole term of $\Ahat_{qq,Q}^{\sf TR, (3)}$,
\begin{eqnarray} 
{\gamma}_{qq}^{\sf TR, (2)} &=& n_f^2 T_F^2 C_F \Biggl\{
{\frac {128}{9}}\,S_3
-{\frac {640}{27}}\,S_2
-{\frac {128}{27}}\,S_1
+{\frac {8}{9}}\,{\frac {( 17\,{N}^{2}+17\,N-8 )}  {N ( 1+N ) }}
\Biggr\}\label{gammaTrans}~.
\end{eqnarray}  
The results for the anomalous dimensions constitute a first independent check of the result obtained in 
\cite{Gracey:2003yrxGracey:2003mrxGracey:2006zrxGracey:2006ah,Blumlein:2009rg}. Again for this color factor 
the vector- and tensor operators lead to the same structures in the harmonic sums.
\section{Conclusions}
\label{sec:concl}

\vspace{2mm}
\noindent
We calculated the $O(n_f)$ contributions to the massive operator matrix elements
at $O(\alpha_s^3)$ contributing to the heavy flavor Wilson coefficients of the
deep-inelastic structure function  $F_2(x,Q^2)$ and to transversity in the
asymptotic region for general values of the Mellin variable $N$ in the 
$\overline{\rm MS}$--scheme. Two of the 3--loop 
OMEs, $A_{qq,Q}^{\sf PS, (3)}$ and $A_{qg,Q}^{\sf (3)}$, are known completely now.
The Feynman diagrams contributing are characterized by one massive and (at least) one 
massless fermion line, with both bubble- and ladder-topologies. The local operator 
insertions are linked to two fermion lines and a number of gluon lines.
The computation of the Feynman parameter integrals has been performed directly 
by representing the integrals as nested sums over generalized hypergeometric functions,
which result into multiple nested sums over products of hypergeometric expressions
and harmonic sums. The sums have been solved by applying modern summation technologies
in difference and product fields. Although in intermediary results in part of the calculation  
generalizations of harmonic sums occurred, the final results can be represented in terms of
rational expressions of the Mellin variable $N$ and of harmonic sums of maximal weight {\sf w=4}.
The harmonic sums contributing show the same structural pattern as being observed in all other massless
2-- and 3--loop calculations. Applying also the structural relations, six harmonic sums span the results.
The small- and large $x$ behaviour of the constant parts of the OMEs has been investigated. In both cases
a less singular behaviour than for the  massless Wilson coefficients is observed. The OMEs
$A_{qq,Q}^{\sf PS, (3)}$ and $A_{qg,Q}^{\sf (3)}$, being completed, do not contain the constant
$\zeta_2$ after renormalization. All results were compared to the fixed moments given in 
\cite{Bierenbaum:2009mv}. We mention that the present calculation is technically very different from that 
of computing fixed moments carried out previously. From the single pole parts in the dimensional 
parameter $\ep$ of the unrenormalized OMEs one may derive the 
respective contributions to the 3--loop anomalous dimensions, which are obtained in three cases as a 
first independent recalculation, using a different method. We confirm the results in the
literature, both in the deep-inelastic case and for transversity.

\newpage
\begin{appendix}
\section{\bf Examples for sums occurring in the calculation}
\label{app1}
\renewcommand{\theequation}{\thesection.\arabic{equation}}
\setcounter{equation}{0}

\vspace{2mm}
\noindent
In the present calculation numerous single-- to triple finite and infinite
sums of an extension of the hypergeometric type had to be calculated. For these sums, depending
on various summation parameters, $n_i$, the ratio of the summands, except the part containing harmonic
sums,
\begin{eqnarray}
\frac{a(...,n_i +1, ...)}
     {a(...,n_i, ...)},~~~~~\forall i
\end{eqnarray}
is a rational function in all variables $n_i$. Sums of this type can be represented
by basic sums of a certain type, which are transcendental to each other and form
sum-- and product--fields, cf.~\cite{SIGMA} and
references therein.
The general form of these sums is
\begin{eqnarray}
\sum_{k_1 = 1}^{N_1(N)}
\sum_{k_2 = 1}^{N_2(k_1,N)}
\sum_{k_3 = 1}^{N_3(k_1,k_2,N)} P(S_{\vec{a_1}}(\tilde{s}_1(k_i,N)),\dots,S_{\vec{a_4}}(\tilde{s}_4(k_i,N))) \Gamma\left[
\begin{array}{l}
s_1(k_i,N), ..., s_p(k_i,N)\\
s_{p+1}(k_i,N), ..., s_{p+q}(k_i,N) \end{array}\right]~,
\end{eqnarray}
with $P(x_1,x_2,x_3,x_4)$ a polynomial from 
$\mathbb{Q}(k_1,k_2,k_3,k_4,N)[x_1,x_2,x_3,x_4]$, with $\tilde{s}_1(k_i,N),$ $
\dots,\tilde{s}_4(k_i,N)$ and $s_1(k_i,N),\dots,
s_{p+q}(k_i,N)$ for some $p, q \in \mathbb{N}$ being integer linear in $k_1,k_2,k_3,N$, with $\vec{a_l}$ an index set, and with the upper bounds $N_1(N), N_2(k_1,N),N_3(k_1,k_2)$ being either $\infty$ or being integer linear in its arguments. The generalized $\Gamma$--function, cf.~\cite{HYPER}, usually includes both
Beta--functions and binomials.

In the present calculation one faces more complicated sums than occurring in earlier two--loop
calculations up to $O(\varepsilon)$, \cite{TOL2}.
Partly they may reach higher weight than appearing in the final result.
In the following we present a few examples.

\begin{eqnarray}
&& \sum_{j_1=1}^{N-2} \sum_{n=1}^{\infty }
(-1)^{j_1} B(n,N-j_1) \binom{N-2}{j_1} \frac{S_2(-j_1+n+N)}{n^2 (j_1-N-2)} =
\N\\
&&
\Biggl\{(-1)^N \frac{6-23 N+9 N^2+2 N^3}{2 (N-1)^2 N^2 (1+N) (2+N)}
+\Biggl[\frac{1}{N+2}-\frac{27 (-1)^N}{(N-1) N (N+1) (N+2)}\Biggr] S_1
\N\\
&&
-\frac{1}{N (N+2)}\Biggr\} S_2^2
\N\\
&&
+\Biggl[\frac{1}{N+2}-\frac{48 (-1)^N}{(N-1) N (N+1) (N+2)}\Biggr] S_3 S_2-\frac{2 S_{-2}^2}{N (N+2)}
\N\\
&&
+\Biggl\{-(-1)^N \frac{7 \big(12+6 N-37 N^2+6 N^3+N^4\big)}{20 (-1+N)^2 N^2 (1+N) (2+N)}
+\Biggl[-(-1)^N \frac{21}{5 (-1+N) N (1+N) (2+N)}
\N\\
&&
-\frac{7}{10 (N+2)}\Biggr] S_1+\frac{7}{10 N (N+2)}\Biggr\} \zeta_2^2+\Biggl\{(-1)^N \frac{6-23 N+9
N^2+2 N^3}{2 (-1+N)^2 N^2 (1+N) (2+N)}
\N\\
&&
+\Biggl[\frac{3 (-1)^N}{(N-1) N (N+1) (N+2)}+\frac{3}{N+2}\Biggr] S_1-\frac{3}{N (N+2)}\Biggr\} S_4
\N\\
&&
+ \Biggl[\frac{3}{N+2}-\frac{18 (-1)^N}{(N-1) N (N+1) (N+2)}\Biggr] S_5
+ \Biggl[\frac{2 S_{-2}^2}{N+2}+\frac{(-1)^N (3 N-1)}{(N-1)^3 N^3}\Biggr] S_1
\N\\
&&
+\frac{2}{2+N} S_{-2} S_{-3} + \Biggl\{-(-1)^N \frac{3 \big(12-6 N-14 N^2+7 N^3+12
N^4+N^5\big)}{(-1+N)^3 N^3 (1+N) (2+N)}
\N
\end{eqnarray}
\begin{eqnarray}
&&
+(-1)^N \frac{9 S_1^2}{(-1+N) N (1+N) (2+N)}+(-1)^N
\frac{3 \big(6-23 N+9 N^2+2 N^3\big)}{(-1+N)^2 N^2 (1+N) (2+N)} S_1
\N\\
&&
+\Biggl[\frac{3 (-1)^N}{(N-1) N (N+1) (N+2)}-\frac{1}{N+2}\Biggr] S_2\Biggr\} S_{2,1}
\N\\
&&
+\Biggl[(-1)^N \frac{2 \big(12-37 N+9 N^2+4 N^3\big)}{(-1+N)^2 N^2 (1+N) (2+N)}+(-1)^N \frac{24}{(-1+N)
N (1+N) (2+N)} S_1\Biggr] S_{3,1}
\N\\
&&
+\frac{2 S_{3,2}}{N+2}+\Biggl[-\frac{12 (-1)^N}{(N-1) N (N+1) (N+2)}-\frac{3}{N+2}\Biggr]
S_{4,1}
\N\\
&&
- \frac{2 S_{-2} S_{-2,1}}{2+N}
+ \frac{4 S_{-3,-2}}{N+2}+\Biggl[-(-1)^N \frac{2 \big(6-23 N+9 N^2+2 N^3\big)}{(-1+N)^2 N^2 (1+N)
(2+N)}
\N\\
&&
-(-1)^N \frac{12}{(-1+N) N (1+N) (2+N)} S_1\Biggr] S_{2,1,1}
-\frac{2 S_{-2,1,-2}}{N+2}
\N\\
&&
+(-1)^N \frac{1}{(-1+N) N (1+N) (2+N)} \Biggl[ 42 S_{2,2,1}
- 24 S_{3,1,1} + 54 S_{2,1,1,1} \Biggr]
\N\\
&&
-(-1)^N \frac{30}{(-1+N) N (1+N) (2+N)} S_3 \tilde{S}_1\left(\frac{1}{2}\right) \tilde{S}_1(2)
\N\\
&&
+(-1)^N \frac{30}{(-1+N) N (1+N) (2+N)} S_1 \tilde{S}_1\left(\frac{1}{2}\right) \tilde{S}_3(2)
+\Biggl\{\frac{(-1)^N 2^{N+2}}{(-1+N)^3 N}
\N\\
&&
+\Biggl[-(-1)^N \frac{2 \big(6-12 N+7 N^2+N^3\big)}{(-1+N)^3 N^3}+(-1)^N \frac{6 S_1^2}{(-1+N) N (1+N)
(2+N)}
\N\\
&&
+(-1)^N \frac{2 \big(6-23 N+9 N^2+2 N^3\big)}{(-1+N)^2 N^2 (1+N) (2+N)} S_1\Biggr] \tilde{S}_1(2)
\N\\
&&
+\Biggl[(-1)^N \frac{2 \big(6-23 N+9 N^2+2 N^3\big)}
{(-1+N)^2 N^2 (1+N) (2+N)}+(-1)^N \frac{12}{(-1+N)
N (1+N) (2+N)} S_1\Biggr] \tilde{S}_2(2)
\N\\
&&
+(-1)^N \frac{6}{(-1+N) N (1+N) (2+N)} \tilde{S}_3(2)\Biggr\} \tilde{S}_{1,1}\left(\frac{1}{2},1\right)
\N\\
&&
+\Biggl\{\Biggl[(-1)^N \frac{12 S_1^2}{(-1+N) N (1+N) (2+N)}
+(-1)^N \frac{2 \big(6-23 N+9 N^2+2 N^3\big)}{(-1+N)^2 N^2 (1+N) (2+N)} S_1\Biggr]
\tilde{S}_1\left(\frac{1}{2}\right)
\N\\
&&
+\Biggl[-(-1)^N \frac{2 \big(6-23 N+9 N^2+2 N^3\big)}{(-1+N)^2 N^2 (1+N) (2+N)}
\N\\
&&
-(-1)^N \frac{12}{(-1+N) N (1+N) (2+N)} S_1\Biggr] \tilde{S}_{1,1}\left(\frac{1}{2},1\right)\Biggr\}
\tilde{S}_{1,1}(2,1)
\N\\
&&
+\Biggl[-(-1)^N \frac{2 \big(6-12 N+7 N^2+N^3\big)}{(-1+N)^3 N^3}
+(-1)^N \frac{6 S_1^2}{(-1+N) N (1+N) (2+N)}
\N\\
&&
+(-1)^N \frac{2 \big(6-23 N+9 N^2+2 N^3\big)}{(-1+N)^2 N^2 (1+N) (2+N)} S_1
\N
\end{eqnarray}
\begin{eqnarray}
&&
+(-1)^N \frac{30}{(-1+N) N (1+N) (2+N)} S_2\Biggr] \tilde{S}_{1,2}\left(\frac{1}{2},2\right)
\N\\
&&
-(-1)^N \frac{6}{(-1+N) N (1+N) (2+N)} \tilde{S}_{1,1}\left(\frac{1}{2},1\right) \tilde{S}_{1,2}(2,1)
\N\\
&&
+\Biggl[(-1)^N \frac{2 \big(6-23 N+9 N^2+2 N^3\big)}{(-1+N)^2 N^2 (1+N) (2+N)}
+(-1)^N \frac{12}{(-1+N) N (1+N) (2+N)} S_1\Biggr]
\N\\ &&
\times \left[
\tilde{S}_{1,3}\left(\frac{1}{2},2\right)
- \tilde{S}_{1,3}\left(2,\frac{1}{2}\right) \right]
\N\\
&&
+ \frac{(-1)^N}{(-1+N) N (1+N) (2+N)} \Biggl\{ 30 \tilde{S}_1\left(\frac{1}{2}\right)
\tilde{S}_{1,3}(2,1)
                                           + 36 \tilde{S}_{1,4}\left(\frac{1}{2},2\right)
\N\\
&&
                                 +  \Biggl[  - 30 S_1 \tilde{S}_1\left(\frac{1}{2}\right)
                                           - 30 \tilde{S}_2\left(\frac{1}{2}\right)
                                           + 30 \tilde{S}_{1,1}\left(\frac{1}{2},1\right)\Biggr]
\tilde{S}_{2,1}(1,2)
\Biggr\}
\N
\\&&
+\Biggl\{\Biggl[(-1)^N \frac{2 \big(6-23 N+9 N^2+2 N^3\big)}{(-1+N)^2 N^2 (1+N) (2+N)}
\N\\
&&
+(-1)^N \frac{24}{(-1+N) N (1+N) (2+N)} S_1\Biggr] 
\tilde{S}_1\left(\frac{1}{2}\right) \Biggr\} \tilde{S}_{2,1}(2,1)
\N\\
&&
+
\frac{(-1)^N}{(-1+N) N (1+N) (2+N)} \Biggl[
- 12 \tilde{S}_{1,1}\left(\frac{1}{2},1\right) \tilde{S}_{2,1}(2,1)
+ 30 \tilde{S}_{2,3}\left(\frac{1}{2},2\right)
\N\\
&&
- 12 \tilde{S}_{2,3}\left(2,\frac{1}{2}\right)
- 18 \tilde{S}_1\left(\frac{1}{2}\right) \tilde{S}_{3,1}(2,1)
+ 30 \tilde{S}_{3,2}\left(\frac{1}{2},2\right)
- 30 \tilde{S}_{4,1}\left(\frac{1}{2},2\right) \Biggr]
\N\\
&&
+\Biggl[(-1)^N \frac{2 \big(6-12 N+7 N^2+N^3\big)}{(-1+N)^3 N^3}
-(-1)^N \frac{6 S_1^2}{(-1+N) N (1+N) (2+N)}
\N
\\&&
-(-1)^N \frac{2 \big(6-23 N+9 N^2+2 N^3\big)}{(-1+N)^2 N^2 (1+N) (2+N)} S_1
\N\\
&&
-(-1)^N \frac{30}{(-1+N) N (1+N) (2+N)} S_2\Biggr] \Biggl[
\tilde{S}_{1,1,1}\left(\frac{1}{2},1,2\right)
+
\tilde{S}_{1,1,1}\left(\frac{1}{2},2,1\right) \Biggr]
\N\\
&&
+\Biggl[-(-1)^N \frac{2 \big(6-23 N+9 N^2+2 N^3\big)}{(-1+N)^2 N^2 (1+N) (2+N)}
\N\\
&&
-(-1)^N \frac{12}{(-1+N) N (1+N) (2+N)} S_1\Biggr] \tilde{S}_1\left(\frac{1}{2}\right)
\tilde{S}_{1,1,1}(1,2,1)
\N\\
&&
+(-1)^N \frac{12}{(-1+N) N (1+N) (2+N)} \tilde{S}_{1,1}\left(\frac{1}{2},1\right)
\tilde{S}_{1,1,1}(2,1,1)
\N\\
&&
+\Biggl[-(-1)^N \frac{2 \big(6-23 N+9 N^2+2 N^3\big)}{(-1+N)^2 N^2 (1+N) (2+N)}
-(-1)^N \frac{12}{(-1+N) N (1+N) (2+N)} S_1\Biggr]
\N\\
&& \times
\Biggl[
     \tilde{S}_{1,1,2}\left(\frac{1}{2},1,2\right)
   - \tilde{S}_{1,1,2}\left(2,\frac{1}{2},1\right)
 - 2 \tilde{S}_{1,1,2}\left(2,1,\frac{1}{2}\right)
\Biggr]
\N
\end{eqnarray}
\begin{eqnarray}
&&
-\frac{(-1)^N}{(-1+N) N (1+N) (2+N)}
\Biggl[
  66 \tilde{S}_{1,1,3}\left(\frac{1}{2},1,2\right)
+ 36 \tilde{S}_{1,1,3}\left(\frac{1}{2},2,1\right)
\N\\
&&
+ 30 \tilde{S}_{1,1,3}\left(1,\frac{1}{2},2\right)
+ 30 \tilde{S}_{1,1,3}\left(1,2,\frac{1}{2}\right)
\Biggr]
\N\\
&&
+\Biggl[-(-1)^N \frac{4 \big(6-23 N+9 N^2+2 N^3\big)}{(-1+N)^2 N^2 (1+N) (2+N)}
         -(-1)^N \frac{24}{(-1+N) N (1+N) (2+N)} S_1\Biggr]
\N\\
&&
 \times \Biggl[ \tilde{S}_{1,2,1}\left(\frac{1}{2},2,1\right)
-\tilde{S}_{1,2,1}\left(2,\frac{1}{2},1\right)
- \frac{1}{2}
 \tilde{S}_{1,2,1}\left(2,1,\frac{1}{2}\right)
\Biggr]
\N\\
&&
-(-1)^N \frac{12}{(-1+N) N (1+N) (2+N)} \tilde{S}_1\left(\frac{1}{2}\right) \tilde{S}_{1,2,1}(1,2,1)
\N\\ &&
-(-1)^N \frac{1}{(-1+N) N (1+N) (2+N)} \Biggl[
  30 \tilde{S}_{1,2,2}\left(\frac{1}{2},1,2\right)
 +36 \tilde{S}_{1,2,2}\left(\frac{1}{2},2,1\right)
\nonumber\\
&&
 +48 \tilde{S}_{1,3,1}\left(\frac{1}{2},2,1\right)
 +30 \tilde{S}_{1,3,1}\left(1,\frac{1}{2},2\right)
+ 30 \tilde{S}_{1,3,1}\left(1,2,\frac{1}{2}\right)
+ 30 \tilde{S}_{2,1,2}\left(\frac{1}{2},2,1\right)
\N\\
&&
+ 30 \tilde{S}_{2,1,2}\left(1,\frac{1}{2},2\right)
- 12 \tilde{S}_{2,1,2}\left(2,\frac{1}{2},1\right)
- 24 \tilde{S}_{2,1,2}\left(2,1,\frac{1}{2}\right)
+ 30 \tilde{S}_{2,2,1}\left(1,2,\frac{1}{2}\right)
\N\\
&&
- 24 \tilde{S}_{2,2,1}\left(2,\frac{1}{2},1\right)
- 12 \tilde{S}_{2,2,1}\left(2,1,\frac{1}{2}\right)
+ 30 \tilde{S}_{3,1,1}\left(\frac{1}{2},1,2\right)
+ 30 \tilde{S}_{3,1,1}\left(\frac{1}{2},2,1\right) \Biggr]
\N\\ &&
+\Biggl[(-1)^N \frac{2 \big(6-23 N+9 N^2+2 N^3\big)}{(-1+N)^2 N^2 (1+N) (2+N)}
\N\\ &&
+(-1)^N \frac{12}{(-1+N) N (1+N) (2+N)} S_1\Biggr]
\Biggl[
   \tilde{S}_{1,1,1,1}\left(\frac{1}{2},1,2,1\right)
\N\\ &&
-2 \tilde{S}_{1,1,1,1}\left(2,\frac{1}{2},1,1\right)
-2 \tilde{S}_{1,1,1,1}\left(2,1,\frac{1}{2},1\right)
-2 \tilde{S}_{1,1,1,1}\left(2,1,1,\frac{1}{2}\right)
\Biggr]
\N\\
&&
+\frac{(-1)^N}{(-1+N) N (1+N) (2+N)}
\Biggl[ 66 \tilde{S}_{1,1,1,2}\left(\frac{1}{2},1,2,1\right)
      + 48 \tilde{S}_{1,1,1,2}\left(\frac{1}{2},2,1,1\right)
\N\\
&&
      + 30 \tilde{S}_{1,1,1,2}\left(1,\frac{1}{2},2,1\right)
      + 30 \tilde{S}_{1,1,1,2}\left(1,2,\frac{1}{2},1\right)
      + 30 \tilde{S}_{1,1,2,1}\left(\frac{1}{2},1,1,2\right)
\N\\
&&
      + 12 \tilde{S}_{1,1,2,1}\left(\frac{1}{2},1,2,1\right)
      + 48 \tilde{S}_{1,1,2,1}\left(\frac{1}{2},2,1,1\right)
      + 30 \tilde{S}_{1,1,2,1}\left(1,\frac{1}{2},1,2\right)
\N\\
&&
      + 30 \tilde{S}_{1,1,2,1}\left(1,2,1,\frac{1}{2}\right)
      + 30 \tilde{S}_{1,2,1,1}\left(\frac{1}{2},1,1,2\right)
      + 30 \tilde{S}_{1,2,1,1}\left(\frac{1}{2},1,2,1\right)
\N\\
&&
      + 12 \tilde{S}_{1,2,1,1}\left(\frac{1}{2},2,1,1\right)
      + 30 \tilde{S}_{1,2,1,1}\left(1,1,\frac{1}{2},2\right)
      + 30 \tilde{S}_{1,2,1,1}\left(1,1,2,\frac{1}{2}\right)
\N\\
&&
      + 30 \tilde{S}_{2,1,1,1}\left(1,\frac{1}{2},1,2\right)
      + 30 \tilde{S}_{2,1,1,1}\left(1,\frac{1}{2},2,1\right)
      + 30 \tilde{S}_{2,1,1,1}\left(1,1,\frac{1}{2},2\right)
\N\\
&&
      + 30 \tilde{S}_{2,1,1,1}\left(1,1,2,\frac{1}{2}\right)
      + 30 \tilde{S}_{2,1,1,1}\left(1,2,\frac{1}{2},1\right)
      + 30 \tilde{S}_{2,1,1,1}\left(1,2,1,\frac{1}{2}\right)
\N
\end{eqnarray}
\begin{eqnarray}
&&
- 24 \tilde{S}_{2,1,1,1}\left(2,\frac{1}{2},1,1\right)
- 24 \tilde{S}_{2,1,1,1}\left(2,1,\frac{1}{2},1\right)
- 24 \tilde{S}_{2,1,1,1}\left(2,1,1,\frac{1}{2}\right)
\N\\
&&
- 12 \tilde{S}_{1,1,1,1,1}\left(\frac{1}{2},1,2,1,1\right)
- 36 \tilde{S}_{1,1,1,1,1}\left(\frac{1}{2},2,1,1,1\right)
\Biggr]
\N\\
&&
+\Biggl\{(-1)^N \frac{3 S_1^2}{(-1+N) N (1+N) (2+N)}
+(-1)^N \frac{2+9 N-5 N^2}{(-1+N)^2 N (1+N) (2+N)}
\N\\ &&
+(-1)^N \frac{6-11 N+2 N^2}{(-1+N)^2 N^2 (2+N)} S_1
\N\\ &&
+\Biggl[\frac{3 (-1)^N}{(N-1) N (N+1) (N+2)}
+\frac{1}{N+2}\Biggr] S_2\Biggr\} \zeta_3
\N\\ &&
+\zeta_2 \Biggl\{
(-1)^N \frac{9 S_1^2}{2 (-1+N) N (1+N) (2+N)}
\N\\ &&
+(-1)^N \frac{2+3 N-2^{2+N} N-2 N^2-3 \cdot 2^{1+N} N^2+6 N^3-2^{1+N} N^3-3 N^4}{(-1+N)^3 N^2 (1+N)
(2+N)}
\N\\ &&
+\Biggl\{
(-1)^N \frac{-12+N+27 N^2-4 N^3}{2 (-1+N)^2 N^2 (1+N) (2+N)}
\N\\ &&
-\Biggl[\frac{1}{N+2}+\frac{6 (-1)^N}{(N-1) N (N+1) (N+2)}\Biggr] S_1
\N\\ &&
+\frac{1}{N (N+2)}\Biggr\} S_2
+\Biggl[-\frac{6 (-1)^N}{(N-1) N (N+1) (N+2)}
-\frac{2}{N+2}\Biggr] S_3
\N\\ &&
-\frac{2 S_{-2}}{N (N+2)}
+ \Biggl[(-1)^N \frac{-6+3 N+18 N^2-20 N^3-3 N^4+2 N^5}{(-1+N)^3 N^3 (1+N) (2+N)}
\N\\ &&
+\frac{2 S_{-2}}{N+2}\Biggr]
 S_1
+\frac{3 S_{-3}}{N+2}
+(-1)^N \frac{12}{(-1+N) N (1+N) (2+N)} S_{2,1}
\N\\ &&
-\frac{2 S_{-2,1}}{N+2}+\bigg[-(-1)^N \frac{3 S_1^2}{(-1+N) N (1+N) (2+N)}
\N\\ &&
+(-1)^N \frac{6-12 N+7 N^2+N^3}{(-1+N)^3 N^3}+(-1)^N \frac{-6+23 N-9 N^2-2 N^3}{(-1+N)^2 N^2 (1+N) (2+N)}
S_1\bigg] \tilde{S}_1(2)
\N\\
&&
+\bigg[(-1)^N \frac{-6+23 N-9 N^2-2 N^3}{(-1+N)^2 N^2 (1+N) (2+N)}
\N\\ &&
-(-1)^N \frac{6}{(-1+N) N (1+N) (2+N)} S_1\bigg] \tilde{S}_2(2)
\N\\ &&
+\bigg[(-1)^N \frac{6-23 N+9 N^2+2 N^3}{(-1+N)^2 N^2 (1+N) (2+N)}
\N\\ &&
+(-1)^N \frac{6}{(-1+N) N (1+N) (2+N)} S_1
\bigg]
\tilde{S}_{1,1}(2,1)
\N\\ &&
-(-1)^N \frac{3}{(-1+N) N (1+N) (2+N)} \tilde{S}_3(2)
\N\\
&&
+\frac{(-1)^N}{(-1+N) N (1+N) (2+N)}\Biggl[
  3 \tilde{S}_{1,2}(2,1)
-15 \tilde{S}_{2,1}(1,2)
+ 6 \tilde{S}_{2,1}(2,1)
- 6 \tilde{S}_{1,1,1}(2,1,1) \Biggr]
\N
\end{eqnarray}
\begin{eqnarray}
&&
+\Biggl[\frac{3}{N+2}-\frac{18 (-1)^N}{(N-1) N (N+1) (N+2)}\Biggr] \zeta_3
\Biggr\}
\N\\ &&
+\Biggl[\frac{27 (-1)^N}{(N-1) N (N+1) (N+2)}-\frac{9}{2 (N+2)}\Biggr] \zeta_5
\N\\ &&
- 30 (-1)^N \frac{1}{(N-1) N (N+1)(N+2)} \tilde{S}_1\left(\frac{1}{2}\right) \tilde{S}_{1,1,2}(2,1,1)~,
\label{SUM:main}
\\
\N\\
{\rm with} \N\\
&& B(a,b) = \frac{\Gamma(a) \Gamma(b)}{\Gamma(a+b)}~.\\
\N\\
 &&\sum_{j=1}^{N-2} \sum_{j_1=1}^{-j+N-2} \sum_{n=1}^{\infty } \frac{(-1)^{j_1} j B(j,n) \binom{-j+N-2}{j_1} 
S_1(j) S_1(n)}{(j+n) (j+n+1) (j+n+2) (j+n+3) (j_1-N-2)} =
\N\\
&&
 (-1)^N \frac{-1+2 N-3 N^2}{N^3 (1+N) (2+N)}
+\frac{-16+12 N+10 N^2-17 N^3-31 N^4}{8 N^3 (1+N) (2+N)}
\N\\
&&
+
\Bigg[(-1)^N\Big[
-\frac{1}{N (2+N)} S_1
+\frac{1}{(N+1) (N+2)}
\Big]
+\frac{1}{4 (N+2)}\Bigg] S_3-\frac{S_4}{2 (N+2)}
\N\\
&&
-(-1)^N \frac{4}{(1+N) (2+N)} S_{-2}
+S_2 \Biggl\{\frac{-8-2 N+N^2+5 N^3+5 N^4}{4 N^3 (1+N) (2+N)}
\N\\
&&
+\Bigg[\frac{(-1)^N (N-1)}{N^2 (N+2)}
+\frac{1}{2 N^2 (N+1)}\Bigg] S_1
+(-1)^N \frac{2}{N (2+N)} S_{-2}
\N\\
&&
-\frac{2 (-1)^N}{(N+1) (N+2)}\Biggr\}
+(-1)^N \frac{1+2 N^2}{N^2 (1+N) (2+N)} S_{-3}
+S_1 \Biggl\{
\frac{2+N+N^2}{2 N^2 (1+N) (2+N)}
\N\\
&&
+(-1)^N \Big[
\frac{2 (-1+N)}{N^2 (2+N)} S_{-2}
-\frac{3}{N (2+N)} S_{-3}\Big] \Biggr\}
+ \frac{3 (-1)^N}{N (2+N)} S_{-4}
\N\\
&&
+\Big[\frac{(-1)^N (1-N)}{N^2 (N+2)}
+\frac{1}{2 (N+2)}\Big] S_{2,1}
-(-1)^N \frac{2}{N (2+N)} S_{2,-2}
\N\\
&&
+\Big[\frac{(-1)^N}{N (N+2)}
+\frac{1}{2 (N+2)}\Big] S_{3,1}
+\Big[(-1)^N \frac{2}{N (2+N)} S_1
\N\\
&&
-(-1)^N\frac{2 \big(-1+2 N^2\big)}{N^2 (1+N) (2+N)}\Big] S_{-2,1}
+(-1)^N\Big[
\frac{2}{N (2+N)} S_{-3,1}
\N\\
&&
-\frac{4}{N (2+N)} S_{-2,1,1}
\Big]
+\Big[\frac{16+4 N-2 N^2+N^3+N^4}{8 N^3 (1+N) (2+N)}
-\frac{S_1}{2 N^2 (N+1)}
\N\\
&&
+\frac{(-1)^N}{N^3 (N+1) (N+2)}\Big] \zeta_2~,
\\
\N\\
&& \sum_{j=1}^{N-2} \sum_{j_1=1}^{-j+N-2} \sum_{n=1}^{\infty } \frac{(-1)^{j_1} j B(j,n) \binom{-j+N-2}{j_1} S_2(n)}{(j+n) (j+n+1) (j+n+2) (j+n+3) (j_1-N-2)} =
\N\\
&&
-\frac{S_2^2}{4 (N+2)}
+\Bigg[\frac{(-1)^N S_1^2}{2 N (N+2)}
+\Big[-\frac{(-1)^N}{N (N+2)}
-\frac{1}{2 N^2 (N+1)}\Big] S_1
\N
\end{eqnarray}
\begin{eqnarray}
&&
-\frac{N^2+1}{2 N^2 (N+1)}
+(-1)^N\Big[ \frac{1}{N (2+N)} S_{-2}
+\frac{1}{(N+1) (N+2)}\Big] \Bigg]S_2
\N
\\
&&
+(-1)^N \frac{8-19 N+24 N^2}{8 (-1+N) N^2 (1+N) (2+N)}
+\frac{-48-24 N+71 N^2+95 N^3}{48 N^2 (1+N) (2+N)}
\N\\
&&
-\frac{S_3}{2 N^2 (N+1)}
+\Bigg[-\frac{(-1)^N}{2 N (N+2)}
-\frac{1}{4 (N+2)}\Bigg] S_4
\N\\
&&
+(-1)^N \Bigg[\frac{2}{(1+N) (2+N)} S_{-2}
+\frac{1}{N (2+N)} S_1^2 S_{-2}
+\frac{1}{N (2+N)} S_{-3}
\N
\\
&&
+S_1 \Big[- \frac{2}{N (2+N)} S_{-2}
-\frac{1}{N (2+N)} S_{-3}\Big]
\N\\
&&
-\frac{1}{N (2+N)} S_{-4}
\Bigg]
+\Bigg[-(-1)^N \frac{1}{N (2+N)} S_1
+\frac{1}{2 N^2 (N+1)}
\N\\
&&
+\frac{(-1)^N}{N (N+2)}\Bigg] S_{2,1}
+\Bigg[\frac{2 (-1)^N}{N (N+2)}
-(-1)^N \frac{2}{N (2+N)} S_1\Bigg] S_{-2,1}
\N\\
&&
+(-1)^N
\Bigg[
\frac{1}{N (2+N)} S_{-3,1}
+\frac{1}{N (2+N)} S_{2,1,1}
+\frac{2}{N (2+N)} S_{-2,1,1}
\Bigg]
\N\\
&&
+\Bigg[(-1)^N \frac{-2+N-2 N^2}{2 (-1+N) N^2 (1+N) (2+N)}
+\frac{2+N-N^2-2 N^3}{2 N^2 (1+N) (2+N)}
+\frac{S_1}{2 N^2 (N+1)}
\N\\
&&
+\Big[\frac{1}{2 (N+2)}
-\frac{(-1)^N}{N (N+2)}\Big] S_2
-(-1)^N \frac{2}{N (2+N)} S_{-2}
\Bigg] \zeta_2
\N\\
&&
+\Biggl[\frac{-12-6 N+N^2+N^3}{12 N^2 (1+N) (2+N)}
+\frac{(-1)^N}{(N-1) N^2 (N+1) (N+2)} \Biggr]\zeta_3~,
\\
\N\\
&& \sum_{j=1}^{N-2} \sum_{j_1=1}^{-j+N-2} \frac{(-1)^{j_1} \binom{-j+N-2}{j_1} 
S_1(j) S_2(-j_1+N)}{(j+2) (j_1-N-2)} =
\N\\
&&
\Bigg[\frac{(-1)^N}{2 (N+1) (N+2)}
-\frac{1}{2 (N+2)}\Bigg] S_2^2
+\Bigg[\frac{S_1^2}{2 (N+2)}
\N\\
&&
+\Biggl[\frac{-3-3 N-N^2}{(1+N)^2 (2+N)^2}
+(-1)^N \frac{-4-5 N-3 N^2-N^3}{N (1+N)^2 (2+N)^2}\Bigg] S_1
\N\\
&&
+(-1)^N \frac{8+28 N+37 N^2-42 N^4-38 N^5-14 N^6-2 N^7}{2 N^2 (1+N)^3 (2+N)^3}
\N\\
&&
+\frac{8+28 N+49 N^2+39 N^3+6 N^4-10 N^5-6 N^6-N^7}{N^2 (1+N)^3 (2+N)^3}\Biggr] S_2
\N
\\
&&
+(-1)^N \Bigg[-\frac{2}{(1+N) (2+N)} S_{-2} S_2
+\frac{\big(-8-28 N-27 N^2-8 N^3\big) S_1^2}{2 N^2 (1+N)^3 (2+N)^3}\Bigg]
\N\\
&&
+\frac{4+5 N+3 N^2+N^3}{N (1+N)^3 (2+N)^2}
+(-1)^N \frac{6+2 N-8 N^2-6 N^3-N^4}{(1+N)^3 (2+N)^3}
\N
\end{eqnarray}
\begin{eqnarray}
&&
+(-1)^N\Bigg[ \frac{-4-3 N+2 N^2+3 N^3+N^4}{N (1+N)^2 (2+N)^2}
+ \frac{1}{(1+N) (2+N)} S_1\Bigg] S_3
\N\\
&&
+(-1)^N \Bigg[\frac{3}{2 (1+N) (2+N)} S_4
- \frac{2 \big(-4+2 N^2+N^3\big)}{N^2 (2+N)^2} S_{-2}
\N\\
&&
+ \frac{3 \big(-4-3 N+2 N^2+3 N^3+N^4\big)}{N (1+N)^2 (2+N)^2} S_{-3}
\Bigg]
+S_1 \Bigg[(-1)^N \frac{16+8 N-4 N^2-N^3}{N^3 (2+N)^3}
\N
\\
&&
-(-1)^N \frac{2 \big(4+5 N+3 N^2+N^3\big)}{N (1+N)^2 (2+N)^2} S_{-2}
+(-1)^N \frac{3}{(1+N) (2+N)} S_{-3}
\N\\
&&
+\frac{1}{(N+1)^2 (N+2)}\Bigg]
+ (-1)^N \Bigg[
+\frac{2}{(1+N) (2+N)} S_{-4}
+\frac{4}{(1+N) (2+N)} S_{2,-2}
\N\\
&&
+\Big[- \frac{2 \big(-4-3 N+2 N^2+3 N^3+N^4\big)}{N (1+N)^2 (2+N)^2}
- \frac{2}{(1+N) (2+N)} S_1\Big] S_{-2,1}
\N\\
&&
-\frac{6}{(1+N) (2+N)} S_{-3,1}
+\frac{4}{(1+N) (2+N)} S_{-2,1,1}
\Bigg]
\end{eqnarray}

In the above examples also so-called generalized harmonic sums occur~\cite{GHSUM1,GHSUM2}. They
obey the following recursive definition~:
\begin{eqnarray}
\label{eq:HSUM4}
\widetilde{S}_{m_1,...}(x_1,...; N) &=& \sum_{i_1=1}^N \frac{x_1^{i_1}}{i_1^{m_1}}
\sum_{i_2=1}^{i_1-1} \frac{x_2^{i_2}}{i_2^{m_2}} \widetilde{S}_{m_3,...}(x_3,...; i_2)
\nonumber\\ & & +
\widetilde{S}_{m_1+m_2,m_3,...}( x_1 \cdot x_2,x_3,...; N)~.
\end{eqnarray}
The sums
$\widetilde{S}$ may be reduced to nested harmonic sums for $x_i \in \{-1, 1\}$. In the present calculation the values
of $x_i$ extend to $\{-1/2, 1/2, -2, 2\}$. These sums occur in ladder like structures,
cf.~\cite{Vermaseren:2005qc,Ablinger:2010ha},
but may also emerge if contributions to 3--loop Feynman diagrams, containing a 2-point insertion,
are separated into various terms. They were even observed in case of the more complicated massive 2-loop
graphs~\cite{TOL2} if large expressions are arbitrarily separated. In part of the sums terms
$\propto$ $2^N$, which lead to an exponential growth in the large $N$ limit, occur. However, all these
contributions cancel for each individual diagram. In the present case
the weight of these sums can reach {\sf w = 5} intermediary, depending on the $\varepsilon$--structure
of the contribution, although only {\sf w = 4} sums will emerge in the final results.
Examples for these sums are~:
\begin{eqnarray}
\label{eq:HSUM5}
&&\widetilde{S}_{1}(1/2,N),~~
\widetilde{S}_2(-2;N),~~
\widetilde{S}_{2,1}(-1,2;N),~~
\widetilde{S}_{3,1}(-2,-1/2;N),~~
\nonumber\\ &&
\widetilde{S}_{1,1,1,2}(-1,1/2,2,-1;N),~~
\widetilde{S}_{2,3}(-2,-1/2;N),~~
\nonumber\\&&
\widetilde{S}_{2,2,1}(-1,-1/2,2;N),~~{\rm etc.}
\end{eqnarray}
The algebraic and structural relations for these sums are worked out in
Ref.~\cite{GHSUM2}. Similar to the case of harmonic sums, corresponding basis representations
are obtained. They allow to simplify involved structures as of Eq.~(\ref{SUM:main}) and
finally lead to the reduction of the results for the individual diagrams to a representation
just in terms of nested harmonic sums. The nested sums emerging in this work,  which were not given before in
Refs.~\cite{TOL2} and those being closer related to the structure of harmonic sums \cite{HSUM2},
are of the type illustrated above. The latter have been calculated using C.~Schneider's
packages {\tt Sigma}~\cite{SIGMA}, {\tt EvaluateInfiniteSums}
\cite{EIS} and J.~Ablinger's package {\tt HarmonicSums} \cite{HARMS}.
\end{appendix}

\vspace*{5mm}\noindent
{\bf Acknowledgment.}\\
This work has been supported in part by SFB-TR/9, the EU TMR network HEPTOOLS,
and Austrian Science Fund (FWF) grants P20162-N18 and P20347-N18.

\newpage


\begin{thebibliography}{99}
%
\bibitem{NLO}
  E.~Laenen, S.~Riemersma, J.~Smith, W.L. van Neerven,
  Nucl.\ Phys.\  {\bf B392 } (1993)  162;
229.\\
  S.~Riemersma, J.~Smith, W.~L.~van Neerven,
  Phys.\ Lett.\  {\bf B347 } (1995)  143,
  [hep-ph/9411431];\\
  Precise representations in Mellin space were derived in~:
  S.~I.~Alekhin and J.~Bl\"umlein,
  Phys.\ Lett.\  B {\bf 594} (2004) 299,
  [arXiv:hep-ph/0404034].
%
\bibitem{H1Z:2009wt}
  F.~D.~Aaron {\it et al.} [ H1 and ZEUS Collaborations ],
  JHEP {\bf 1001 } (2010)  109,
  [arXiv:0911.0884 [hep-ex]].
%
\bibitem{PDF}
  S.~Alekhin, J.~Bl{\"u}mlein and S.~Moch,
  arXiv:1007.3657 [hep-ph], and in preparation\\
  S.~Alekhin, J.~Bl\"umlein, S.~Klein and S.~Moch,
  Phys.\ Rev.\  D {\bf 81} (2010) 014032,
  [arXiv:0908.2766 [hep-ph]];\\
  J.~Bl{\"u}mlein, H.~B{\"o}ttcher and A.~Guffanti,
  Nucl.\ Phys.\  B {\bf 774} (2007) 182,
  [arXiv:hep-ph/0607200];
  Nucl.\ Phys.\ Proc.\ Suppl.\  {\bf 135} (2004) 152,
  [arXiv:hep-ph/0407089];\\
  M.~Gl{\"u}ck, E.~Reya and C.~Schuck,
  Nucl.\ Phys.\  B {\bf 754} (2006) 178,
  [arXiv:hep-ph/0604116];\\
  P.~Jimenez-Delgado and E.~Reya,
  Phys.\ Rev.\  D {\bf 79} (2009) 074023,
  [arXiv:0810.4274 [hep-ph]];\\
  A.~D.~Martin, W.~J.~Stirling, R.~S.~Thorne and G.~Watt,
  Eur.\ Phys.\ J.\  C {\bf 63} (2009) 189,
  [arXiv:0901.0002 [hep-ph]].
%
\bibitem{EXP}
  F.~D.~Aaron {\it et al.}  [H1 Collaboration],
  arXiv:1008.1731 [hep-ex];
  Eur.\ Phys.\ J.\  C {\bf 65} (2010) 89
  [arXiv:0907.2643 [hep-ex]];\\
  A.~Aktas {\it et al.}  [H1 Collaboration],
  Eur.\ Phys.\ J.\  C {\bf 47} (2006) 597
  [arXiv:hep-ex/0605016];
  Eur.\ Phys.\ J.\  C {\bf 41} (2005) 453
  [arXiv:hep-ex/0502010];\\
  S.~Chekanov {\it et al.}  [ZEUS Collaboration],
  Eur.\ Phys.\ J.\  C {\bf 65} (2010) 65
  [arXiv:0904.3487 [hep-ex]];
  JHEP {\bf 0902} (2009) 032
  [arXiv:0811.0894 [hep-ex]].
  Phys.\ Lett.\  B {\bf 599} (2004) 173
  [arXiv:hep-ex/0405069];\\
  H.~Abramowicz {\it et al.}  [ZEUS collaboration],
  arXiv:1005.3396 [hep-ex].
%
\bibitem{HERALHC}
  M.~Dittmar, S.~Forte, A.~Glazov {\it et al.},
  [hep-ph/0511119];\\
  S.~Alekhin, G.~Altarelli, N.~Amapane {\it et al.},
  [hep-ph/0601012],[hep-ph/0601013];\\
  Z.~J.~Ajaltouni, S.~Albino, G.~Altarelli {\it et al.},
  [arXiv:0903.3861 [hep-ph]].
%
\bibitem{Blumlein:2008kz}
  J.~Bl\"umlein, H.~B\"ottcher,
  Phys.\ Lett.\  {\bf B662 } (2008)  336-340.
  [arXiv:0802.0408 [hep-ph]].
%
\bibitem{Presti:2010pd}
  N.~A.~L.~Presti, H.~Kawamura, S.~Moch and A.~Vogt,
  arXiv:1008.0951 [hep-ph].
%
\bibitem{Buza:1995ie}
  M.~Buza, Y.~Matiounine, J.~Smith, R.~Migneron and W.~L.~van Neerven,
  Nucl.\ Phys.\  B {\bf 472} (1996) 611,
  [hep-ph/9601302].
%
\bibitem{Vermaseren:2005qc}
  J.~A.~M.~Vermaseren, A.~Vogt and S.~Moch,
  Nucl.\ Phys.\  B {\bf 724} (2005) 3,
  [hep-ph/0504242].
%
\bibitem{Bierenbaum:2009mv}
  I.~Bierenbaum, J.~Bl{\"u}mlein and S.~Klein,
  Nucl.\ Phys.\  B {\bf 820} (2009) 417,
   [hep-ph/0904.3563].
%
\bibitem{Blumlein:2009rg}
  J.~Bl\"umlein, S.~Klein and B.~T\"odtli,
  Phys.\ Rev.\  {\bf D 80} (2009) 094010,
  [arXiv:0909.1547 [hep-ph]].
%
\bibitem{Steinhauser:2000ry}
  M.~Steinhauser,
  Comput.\ Phys.\ Commun.\  {\bf 134} (2001) 335,
  [arXiv:hep-ph/0009029].
%
\bibitem{TOL1a}
  M.~Buza, Y.~Matiounine, J.~Smith and W.~L.~van Neerven,
  Eur.\ Phys.\ J.\  C {\bf 1} (1998) 301,
  [hep-ph/9612398];
%
\bibitem{TOL1b}
  M.~Buza, Y.~Matiounine, J.~Smith and W.~L.~van Neerven,
  Nucl.\ Phys.\  {\bf B485 } (1997)  420,
  [hep-ph/9608342].
%
\bibitem{TOL2}
  I.~Bierenbaum, J.~Bl\"umlein and S.~Klein,
  Nucl.\ Phys.\  B {\bf 780} (2007) 40,
  [hep-ph/0703285];
  Phys.\ Lett.\  B {\bf 648} (2007) 195,
   [hep-ph/0702265];
%
  Phys.\ Lett.\  B {\bf 672} (2009) 401,
  [hep-ph/0901.0669];
  I.~Bierenbaum, J.~Bl\"umlein, S.~Klein and C.~Schneider,
  Nucl.\ Phys.\  B {\bf 803} (2008) 1,
  [hep-ph/0803.0273].
%
\bibitem{Bierenbaum:2010jp}
  I.~Bierenbaum, J.~Bl\"umlein and S.~Klein,
  arXiv:1008.0792 [hep-ph]. 
%
\bibitem{BBKW}
  I.~Bierenbaum, J.~Bl\"umlein, S.~Klein, and F. Wi\ss{}brock, in preparation.
%
\bibitem{ANDIM2a}
  J.~A.~Gracey,
  Phys.\ Lett.\  B {\bf 322} (1994) 141
  [arXiv:hep-ph/9401214].
%
\bibitem{ANDIM2b}
  S.~Moch, J.~A.~M.~Vermaseren and A.~Vogt,
  Nucl.\ Phys.\  B {\bf 688} (2004) 101,
  [hep-ph/0403192].\\
  Nucl.\ Phys.\  B {\bf 691} (2004) 129,
  [hep-ph/0404111].
%
\bibitem{Blumlein:2006mh}
  J.~Bl\"umlein, A.~De Freitas, W.~L.~van Neerven and S.~Klein,
  Nucl.\ Phys.\  B {\bf 755} (2006) 272,
  [hep-ph/0608024].
%
\bibitem{Blumlein:1998sh}
  J.~Bl\"umlein, W.~L.~van Neerven,
  Phys.\ Lett.\  {\bf B450 } (1999)  417-426.
  [hep-ph/9811351].
%
\bibitem{HYPER}
W.~Bailey {\sf Generalized Hypergeometric Series}, (Cambridge University Press,
  Cambridge, 1935), 108~p.;\\
L.~Slater {\sf Generalized Hypergeometric Functions}, (Cambridge University
  Press, Cambridge, 1966), 273~p.
%
\bibitem{Blumlein:2009ta}
  J.~Bl\"umlein,
  Comput.\ Phys.\ Commun.\  {\bf 180} (2009) 2218
  [arXiv:0901.3106 [hep-ph]];
%
\bibitem{Bytev:2009kb}
  V.~V.~Bytev, M.~Y.~Kalmykov and B.~A.~Kniehl,
  Nucl.\ Phys.\  B {\bf 836} (2010) 129
  [arXiv:0904.0214 [hep-th]].
%
\bibitem{SIGMA}
C. Schneider, J. Symbolic Comput. 43 (2008) 611,
\newblock [arXiv:0808.2543v1]; Ann. Comb. 9 (2005) 75;
J. Differ. Equations Appl. 11 (2005) 799; Ann. Comb.  (2010) to appear,
[arXiv:0808.2596]; 
{Proceedings of the Conference on Motives, Quantum Field Theory, and
Pseudodifferential Operators}, To appear in the Mathematics Clay Proceedings,
2010; 
S\'em.~Lothar. Combin. 56 (2007) 1, Article B56b,  Habilitationsschrift JKU Linz (2007)
and references therein;\\
  J.~Ablinger, J.~Bl{\"u}mlein, S.~Klein and C.~Schneider,
  Nucl. Phys. (Proc. Suppl.) {\bf B} (2010), 110, arXiv:1006.4797 [math-ph].
%
\bibitem{HARMS}
J. Ablinger, Diploma Thesis, JKU Linz, February 2009.
%
\bibitem{KARR}
M. Karr, J. ACM. {\bf 28} (1981) 305.
%
\bibitem{AEB}
M. Petkov{\v s}ek, H.S. Wilf, and D. Zeilberger, {\sf A = B} (A. K. Peters, Wellesley, MA, 1996).
%
\bibitem{HSUM1}
  J.~Bl{\"u}mlein and S.~Kurth,
  Phys.\ Rev.\  D {\bf 60} (1999) 014018,
  [arXiv:hep-ph/9810241].
%
\bibitem{HSUM2}
  J.~A.~M.~Vermaseren,
  Int.\ J.\ Mod.\ Phys.\  A {\bf 14} (1999) 2037,
  [arXiv:hep-ph/9806280].
%
\bibitem{GHSUM1}
  S.~Moch, P.~Uwer and S.~Weinzierl,
  J.\ Math.\ Phys.\  {\bf 43} (2002) 3363
  [arXiv:hep-ph/0110083].
%
\bibitem{GHSUM2}
J.~Ablinger, J.~Bl{\"u}mlein and C.~Schneider (2010), in preparation.
%
\bibitem{GHSUM3}
A.~Goncharov, {Math. Res. Lett. {\bf 5} (1998) 497};\\
  J.~M.~Borwein, D.~M.~Bradley, D.~J.~Broadhurst and P.~Lisonek,
  Trans.\ Am.\ Math.\ Soc.\  {\bf 353} (2001) 907
  [arXiv:math/9910045].
%
\bibitem{Gracey:2003yrxGracey:2003mrxGracey:2006zrxGracey:2006ah}
  J.~A.~Gracey,
  Nucl.\ Phys.\  B {\bf 662} (2003) 247
  [arXiv:hep-ph/0304113];
  Nucl.\ Phys.\  B {\bf 667} (2003) 242
  [arXiv:hep-ph/0306163];
  JHEP {\bf 0610} (2006) 040
  [arXiv:hep-ph/0609231];
  Phys.\ Lett.\  B {\bf 643} (2006) 374
  [arXiv:hep-ph/0611071].
%
\bibitem{Ablinger:2010ha}
  J.~Ablinger, I.~Bierenbaum, J.~Bl\"umlein, A Hasselhuhn, S. Klein, C. Schneider, and F.~Wi\ss{}brock,
  Nucl. Phys. (Proc. Suppl.) (2010) 242, [arXiv:1007.0375 [hep-ph]];\\
  J.~Bl\"umlein, A. Hasselhuhn, S. Klein, and C. Schneider, in preparation.
%
\bibitem{Klein:2009ig}
  S.~Klein,
  {\sf Mellin moments of heavy flavor contributions to $F_2(x,Q^2)$ at NNLO}, PhD Thesis,
  TU Dortmund, September 2009,
  arXiv:0910.3101 [hep-ph].
%
\bibitem{FORM}
  J.~A.~M.~Vermaseren,
  arXiv:math-ph/0010025.
%
\bibitem{vanRitbergen:1998pn}
  T.~van Ritbergen, A.~N.~Schellekens, J.~A.~M.~Vermaseren,
  Int.\ J.\ Mod.\ Phys.\  {\bf A14 } (1999)  41-96.
  [hep-ph/9802376].
%
\bibitem{IBP}
J.~Lagrange {\sf Nouvelles recherches sur la nature et la propagation du son},
  Miscellanea Taurinensis, t. II, 1760-61; Oeuvres t. I, p. 263;\\
  K.~G.~Chetyrkin, A.~L.~Kataev, F.~V.~Tkachov,
  Nucl.\ Phys.\  {\bf B174 } (1980)  345;\\
  S.~Laporta,
  Int.\ J.\ Mod.\ Phys.\  {\bf A15 } (2000)  5087-5159.
  [hep-ph/0102033].
%
\bibitem{Blumlein:2009cf}
  J.~Bl\"umlein, D.~J.~Broadhurst and J.~A.~M.~Vermaseren,
  Comput.\ Phys.\ Commun.\  {\bf 181} (2010) 582
  [arXiv:0907.2557 [math-ph]] and references therein.
%
\bibitem{Blumlein:2003gb}
  J.~Bl\"umlein,
  Comput.\ Phys.\ Commun.\  {\bf 159} (2004) 19
  [arXiv:hep-ph/0311046].
%
\bibitem{ANDIM3}
  X.~Artru and M.~Mekhfi,
  Z.\ Phys.\  C {\bf 45} (1990) 669;\\
  J.~Bl\"umlein,
  Eur.\ Phys.\ J.\  C {\bf 20} (2001) 683
  [arXiv:hep-ph/0104099] and references therein;\\
  A.~Hayashigaki, Y.~Kanazawa and Y.~Koike,
  Phys.\ Rev.\  D {\bf 56} (1997) 7350
  [arXiv:hep-ph/9707208];\\
  S.~Kumano and M.~Miyama,
  Phys.\ Rev.\  D {\bf 56} (1997) 2504
  [arXiv:hep-ph/9706420];\\
  W.~Vogelsang,
  Phys.\ Rev.\  D {\bf 57} (1998) 1886
  [arXiv:hep-ph/9706511] and references therein.
%
\bibitem{RATCL}
  V.~Barone, A.~Drago and P.~G.~Ratcliffe,
  Phys.\ Rept.\  {\bf 359} (2002) 1
  [arXiv:hep-ph/0104283].
%
\bibitem{WISSB}
F. Wi\ss{}brock, {\sf $O(\alpha_s^3 T_F^2 N_f)$ Contributions to the Heavy Flavor
Wilson Coefficients of the Structure Function $F_2(x,Q^2)$ at $Q^2 \gg m^2$}, 
Diploma Thesis, Freie Universit\"at Berlin, June 2010.
%
\bibitem{Blumlein:2009tj}
  J.~Bl\"umlein, M.~Kauers, S.~Klein and C.~Schneider,
  Comput.\ Phys.\ Commun.\  {\bf 180} (2009) 2143
  [arXiv:0902.4091 [hep-ph]].
%
\bibitem{BRK}
  J.~Bl\"umlein, V.~Ravindran,
  Nucl.\ Phys.\  {\bf B716 } (2005)  128
  [hep-ph/0501178];
  Nucl.\ Phys.\  {\bf B749 } (2006)  1
  [hep-ph/0604019];\\
  J.~Bl\"umlein, S.~Klein,
  [arXiv:0706.2426 [hep-ph]].
%
\bibitem{STRUC_H}
  J.~Bl\"umlein,
  {\sf Structural Relations of Harmonic Sums and Mellin Transforms at Weight
    w=6}, {Proceedings of the Conference on Motives, Quantum Field Theory, and
Pseudodifferential Operators}, To appear in the Mathematics Clay Proceedings,
2010, 
  arXiv:0901.0837 [math-ph];\\
J. Ablinger, J. Bl\"umlein, and C. Schneider, in preparation.
%
\bibitem{DFLM}
  M.~Diemoz, F.~Ferroni, E.~Longo {\it et al.},
  Z.\ Phys.\  {\bf C39 } (1988)  21.
%
\bibitem{ANCONT}
  J.~Bl\"umlein,
  Comput.\ Phys.\ Commun.\  {\bf 133} (2000) 76
  [arXiv:hep-ph/0003100];\\
  J.~Bl\"umlein and S.~O.~Moch,
  Phys.\ Lett.\  B {\bf 614} (2005) 53
  [arXiv:hep-ph/0503188];\\
  A.~V.~Kotikov, V.~N.~Velizhanin,
  [hep-ph/0501274];\\
  S.~Albino,
  Phys.\ Lett.\  {\bf B674 } (2009)  41-48.
  [arXiv:0902.2148 [hep-ph]].
%
\bibitem{SX}
  J.~Bl\"umlein, A.~Vogt,
  Phys.\ Lett.\  {\bf B370 } (1996)  149
  [hep-ph/9510410];\\
  S.~Catani, F.~Hautmann,
  Nucl.\ Phys.\  {\bf B427 } (1994)  475
  [hep-ph/9405388].
%
\bibitem{Blumlein:2004bs}
  J.~Bl\"umlein, H.~Kawamura,
  Nucl.\ Phys.\  {\bf B708 } (2005)  467
  [hep-ph/0409289].
%
\bibitem{NIELS}
N. Nielsen, {\sf Der Eulersche Dilogarithmus und Seine Verallgemeinerungen}, Nova Acta Leopold.,
Vol. {\bf XC}, Nr. 3, Halle, (1909), pp. 121;\\
S. K\"olbig, Siam J. Math. Anal. {\bf 17} (1986) 1232.
%
\bibitem{ANDIM1} 
  S.~A.~Larin, T.~van Ritbergen and J.~A.~M.~Vermaseren,
  Nucl.\ Phys.\  B {\bf 427} (1994) 41;\\
  S.~A.~Larin, P.~Nogueira, T.~van Ritbergen and J.~A.~M.~Vermaseren,
  Nucl.\ Phys.\  B {\bf 492} (1997) 338,
  [hep-ph/9605317];\\
  A.~Retey and J.~A.~M.~Vermaseren,
  Nucl.\ Phys.\  B {\bf 604} (2001) 281,
  [hep-ph/0007294];\\
  J.~Bl\"umlein and J.~A.~M.~Vermaseren,
  Phys.\ Lett.\  B {\bf 606} (2005) 130,
  [hep-ph/0411111].
%
\bibitem{EIS} 
C. Schneider, in preparation.
\end{thebibliography}
\end{document}